\documentclass[a4paper,fleqn]{cas-dc}

\usepackage[numbers]{natbib}

\newtheorem{definition}{Definition}
\usepackage{enumitem}

\newtheorem{insight}{Insight}
\newcommand{\ignore}[1]{}
\usepackage{caption}
\usepackage{algorithm,algorithmic}

\usepackage{subcaption}
\usepackage{nicematrix}
\usepackage{svg} 
\usepackage{tikz} 

\begin{document}
\let\WriteBookmarks\relax
\def\floatpagepagefraction{1}
\def\textpagefraction{.001}
\newcommand{\PTac}{\text{PTac}}
\newcommand{\PTech}{\text{PTech}}
\shorttitle{Quantifying Malicious Email Sophistication}

\shortauthors{Longtchi et~al.}

\title [mode = title]{Quantifying Psychological Sophistication of Malicious Emails}

\ignore{
\tnotemark[1,2]

 Title footnote 1.
\tnotetext[1]{This document is the results of the research project funded by the National Science Foundation.}

\tnotetext[2]{The second title footnote which is a longer text matter
   to fill through the whole text width and overflow into
   another line in the footnotes area of the first page.}
}

%
\author[1]{Theodore Longtchi}





\credit{TL, RR, DA, CK, and SX conceived the research including methodology. TL and RR collected the emails for analysis. TL, RR, and KG, and EE analyzed the emails with their respective grades. TL and RR analyzed the grades. All authors contributed to the writing of the paper.}

\author[2]{Rosana Monta\~nez Rodriguez}
\author[1]{Kora Gwartney} 
\author[1]{Ekzhin Ear}
\author[3]{David P. Azari}
\author[3]{Christopher P. Kelley}
\author[1]{Shouhuai Xu}

\affiliation[1]{organization={Department of Computer Science, University of Colorado Colorado Springs},
    addressline={1420 Austin Bluffs Pkwy}, 
    city={Colorado Springs},
    postcode={80918}, 
     state={CO},
    country={USA}}



\affiliation[2]{organization={Department of Computer Science, University of Texas at San Antonio},
    city={San Antonio},
    postcode={78249}, 
    state={TX},
    country={USA}}

\affiliation[3]{organization={Department of Behavioral Sciences and Leadership, US Air Force Academy}, 
    city={Colorado Springs},
     citysep={}, 
    postcode={80840}, 
    state={CO},
    country={USA}}

\cortext[cor1]{Corresponding author: Shouhuai Xu ({\tt sxu@uccs.edu})}

\ignore{
\fntext[fn1]{This is the first author footnote. but is common to third
  author as well.}
\fntext[fn2]{Another author footnote, this is a very long footnote and
  it should be a really long footnote. But this footnote is not yet
  sufficiently long enough to make two lines of footnote text.}
}

\nonumnote{The preliminary version of the present paper was published in the Proceedings of SciSec'2023 \cite{rodriguez2023quantifying}.
  }

\begin{abstract}
Malicious emails (including Phishing, Spam, and Scam) are one significant class of cyber social engineering attacks. Despite numerous defenses to counter them, the problem remains largely open. The ineffectiveness of current defenses can be attributed to our superficial understanding of the psychological properties that make these attacks successful. This problem motivates us to investigate the psychological sophistication, or {\it sophistication} for short, of malicious emails. We propose an innovative framework that accommodates two important and complementary aspects of  sophistication, dubbed {\it Psychological Techniques} (PTechs) and {\it Psychological Tactics} (PTacs). We propose metrics and grading rules for human experts to assess the sophistication of malicious emails via the lens of these PTechs and PTacs. To demonstrate the usefulness of the framework, we conduct a case study based on 1,036 malicious emails assessed by four independent graders. Our results show that malicious emails are psychologically sophisticated, while exhibiting both commonalities and different patterns in terms of their PTechs and PTacs. Results also show that previous studies might have focused on dealing with the less proliferated PTechs (e.g., {\sf Persuasion}) and PTacs (e.g., {\tt Reward}), rather than the most proliferated PTechs (e.g., {\sf Attention Grabbing} and {\sf Impersonation}) and  PTacs (e.g., {\tt Fit \& Form} and {\tt Familiarity})) that are identified in this study. We also found among others that social events are widely exploited by attackers in contextualizing their malicious emails.
These findings could be leveraged to guide the design of effective defenses against malicious emails. 

\end{abstract}

\begin{keywords}
cybersecurity metrics \sep  cyber social engineering attacks \sep  malicious emails \sep  psychological factors \sep  psychological tactics \sep  psychological techniques \sep  psychological sophistication
\end{keywords}

\maketitle 

\section{Introduction}
Cyber social engineering attacks 
have become an 
effective weapon for attackers to gain entry points into networks \cite{dbir2021verizon}. The consequences of these attacks have motivated many studies on designing countermeasures against them (e.g., \cite{DBLP:conf/cns/PritomX22,MirIEEEISI2020Landsapce,MirIEEEISI2020MalciousWebsites,nelms2016usenix,jakobsson2007human,goel2017got,nelms2016usenix,vishwanath2011people,anderson2020security,dhamija2006acm}). Despite numerous efforts to stop these attacks, the attacks remain effective. Known defenses are often based on automated identification, typically via machine learning techniques (e.g., recognizing known Phishing email patterns). However, they do not consider psychological factors that can be leveraged to identify new approaches to defend against these attacks \cite{longtchi2024internet}. As suggested in \cite{longtchi2024internet}, the initial step to effectively defending against these attacks is to systematically understand, characterize, and quantify the role of various psychological factors that contribute to their success. In addition, the industry has made many products to counter these attacks, such as FireEye Email Security, Mimecast, Cisco Cloud Email Security, Barracuda Sentinel, Microsoft Defender, and Norton LifeLock. Despite all these efforts, these attacks remain effective. For example, the 2022 Anti-Phishing Working Group (APWG) report \cite{apwg2022report} states that the number of Phishing attacks reported to APWG has quadrupled since early 2020; a more recent report \cite{APWG2023report} states that 
2023 has been the worst year on record so far, with more than 5 million Phishing attacks. 

The preceding discussion inspires us to 
investigate the notion of {\it psychological sophistication}, or {\it sophistication} for short, of malicious emails to pave a way towards designing effective defenses. Our focus is on investigating email sophistication with respect to the psychological elements in malicious emails that attackers employ  to lure victims. That is, our focus is on individual email and its content, which include the sender, the recipient, the day of the year, the subject of the email, and the inner presentation of the email such as the salutation, the body, the signature, and the footer. To our knowledge, these psychological elements have not been systematically studied in the literature. This is true despite the studies that considered the impact of the following non-psychological factors on the success of Phishing attacks: the {time of the day}, the {day of the week}, or the {month of the year} that an email is sent  \cite{ho2019detecting,oest2020sunrise,gascon2018reading}. 

\noindent{\bf Our Contributions}. This paper makes three contributions. First, we initiate a systematic investigation on the psychological sophistication of malicious emails. Specifically, we ask three questions: (i) How can we quantify the psychological sophistication of malicious emails in the real world? (ii) What kinds of psychological sophistication patterns are exhibited by different kinds (or types) of malicious emails in the real world? (iii) How does the sophistication of malicious email evolve over time?

Second, we propose  an innovative and systematic framework to quantify the psychological sophistication of malicious emails. 
The innovation of the framework is to deconstruct low-level and high-level psychological features that influence an email recipient to engage with the malicious content.   The framework deconstructs and compares the content of malicious emails through two lenses:

\begin{itemize}
\item {\it Psychological Techniques (PTechs)}: \ignore{In terms of} At the low level, 
we propose identifying the number of psychologically relevant textual and imagery elements in an email message, dubbed {\it Psychological Techniques} (PTechs) to provide a detailed accounting of the elements employed in implementing an attack. 

\item {\it Psychological Tactics (PTacs)}: At the high level, \ignore{psychological features,} we propose assessing the attacker's overall deliberate thoughtfulness (reflecting attacker's effort) in framing malicious content to influence an email recipient,
dubbed {\it Psychological Tactics} (PTacs), to measure an attacker's effort in exploiting human fallibility.  
\end{itemize}
We propose considering both PTechs and PTacs because they offer complimentary views of an attacker's effort as reflected by the observed elements in malicious emails. Together, they enable a rich exploration of qualitative and quantitative insights into how an attacker frames and implements its attack. As a side-product, the framework could be adapted / adopted to quantify the sophistication of other types of cyber social engineering attacks, such as malicious text messages. 

Third, we demonstrate the usefulness of the framework by applying it to quantify the sophistication of 1,036 malicious emails. This empirical study allows us to draw useful insights, including: (i) previous studies might have focused on dealing with the 
less proliferated PTechs (e.g., {\sf Persuasion}) and PTacs (e.g., {\tt Reward}), rather than the most proliferated PTechs (e.g., {\sf Attention Grabbing} and {\sf Impersonation}) and PTacs (e.g., {\tt Fit \& Form} and {\tt Familiarity}) that are identified in the present study;
(ii) Phishing emails are psychologically more sophisticated than Spam and Scam emails; (iii) PTechs are independently employed by attackers, suggesting no coordination between attackers; (iv) malicious emails that are sophisticated in PTechs are also sophisticated in PTacs.

As a side-product, we present the algorithm for computing the degree of agreement between graders on a given set of objects (i.e., emails in this context) with respect to multiple attributes (i.e., PTechs and PTacs in this context). The idea behind the algorithm was presented in the literature \cite{marzi2024k, schober2018correlation,krippendorff2011computing}, 
but we are not aware of any other algorithmic description in the literature. Thus, the algorithm may be of independent value to the computer science community.
\smallskip

\noindent{\bf Ethical Issue}. In consultation with University of Colorado Colorado Springs Internal Review Board (IRB), this study does not need IRB approval because no subjects are part of the study and the emails are provided by a third party. 

\smallskip

\smallskip

\noindent{\bf{Related Work.}} 
Longtchi et al. \cite{longtchi2024internet} systematize the psychological aspects relevant to cyber social engineering attacks. The present study is inspired by \cite{longtchi2024internet}, and proposes the first framework for quantifying psychological sophistication of malicious emails. The framework can be incorporated into the broader framework presented in \cite{longtchi2024internet} towards ultimately tackling cyber social engineering attacks.

Studies have investigated the use of psychological content in Phishing emails (e.g., \cite{allodi2019need,goel2017got,flores2015investigating,van2019cognitive,nelms2016usenix,ferreira2015analysis}). 
Allodi et al. \cite{allodi2019need} study Phishing and Spear Phishing and defenses, showing that {\sc scarcity} is employed in Phishing emails, but defenses do not account for human-related characteristics. Goel et al. \cite{goel2017got} investigate humans' susceptibility to \textit{deception}. Flores et al. \cite{flores2015investigating} study correlation between personal psychology and demographics in terms of resistance to Phishing, finding that neither age nor gender significantly correlates to Phishing resilience but computer experience does. The other studies examine the psychological content of social engineering messages through the lens of Cialdini's Persuasion Principles and other principles. Heijden and Allodi \cite{van2019cognitive} study the identification of 
persuasion elements in Phishing emails.
Nelms et al. \cite{nelms2016usenix} investigate the 
psychological tactics to encourage users to download malicious applications, by considering 
Whaley's Theory of Deception in addition to Cialdini's Persuasion Principles. Ferreira and Lenzini \cite{ferreira2015analysis} investigate 
the psychological content in Phishing messages based on low-level psychological elements by leveraging Cialdini's Persuasion Principles, Stajano's Scam Principles, and Gragg's psychological triggers. 
In particular, we consider both low-level and high-level psychological features.
De Bona and Paci \cite{de2020real} 
show, among other things, that urgency is more effective than  authority in making employees susceptible to Phishing attacks.
By contrast, we systematically study the impact of PTechs and PTacs on the sophistication of malicious emails.

Studies have investigated the use of psychological content in Spam emails.
Gallo et al. \cite{gallo2024human} design a system 
that can detect persuasive elements in Phishing email, 
such as consistency, {\sf urgency}, {\sc scarcity, liking, social proof, authority}, and {\sc reciprocity}.
Gallo et al. \cite{gallo20212} investigate Spam emails collected in a 2-year span and show 
how Phishing emails exploit cognitive vulnerabilities, such as the use of deceiving words (e.g., \textit{account, suspended} and \textit{verify}) or scamming words (e.g., \textit{donate}, \textit{please}, and \textit{warning}).
Vance et al. \cite{vance2019fog} study the effect of habituation and generalization with respect to warnings and notification signs in an online setting, 
where habituation indicates people's response to repeated stimulation decreases over time and generalization means habituation 
to a stimulus is carried over to other stimuli that are similar in appearance. 
However, these studies do not consider the quantification of psychological sophistication of malicious emails.

To determine how personality traits influence susceptibility to social engineering attack, Uebelacker and Quiel \cite{uebelacker2014social} propose a 
Social Engineering Personality Framework (SEPF) to describe 
the relationships between the personality traits of the Big Five Model and the Cialdini's principles of influence.
The framework aims to support and guide security researchers and practitioners in developing detection, mitigation, and prevention strategies while dealing with human factors in social engineering attacks. While this framework and ours are both geared towards providing insights in social engineering attacks for a better defense system, our paper conducts a case study to test the usefulness of the framework, which lead to some important insights. 
Relevant prior studies (e.g., \cite{wang2021social,frauenstein2020susceptibility,ferreira2015analysis,van2019cognitive,schaab2017social,ferreira2015principles, wang2012research}) that investigate how PTechs are exploited by malicious emails only consider one or very few PTechs. For example, Wang et al. \cite{wang2012research} study the impact of visceral triggers on Phishing susceptibility, but only considering few  visceral triggers (e.g. fear, the stressing of {\sf urgency} to respond, or the implicit use of {\sf impersonation}).
By contrast, we consider the evolution of 7 PTacs and 8 PTechs from 2006 to 2022 with intervals of 5 years (i.e., 2006, 2011, 2036, and 2021 plus 2022) using a large dataset of 1,036 malicious emails.
When compared with these studies, we propose a systematic framework for quantifying the psychological sophistication of malicious emails.
Moreover, our case study additionally considers Spam and Scam emails.

\smallskip
\noindent{\bf Paper Outline}.
Section \ref{sec:preliminaries} describes the psychological concepts used in the paper. Section \ref{Methodology} presents the framework for quantifying sophistication of malicious emails. Section \ref{Experiment Result and Data Analysis} reports a case study on applying the framework to a malicious email dataset. 
Section \ref{sec:limitations} discuses limitations of the study. Section \ref{Conclusion} concludes the paper with research directions.

\section{The Concepts of PTechs and PTacs} \label{sec:preliminaries}

\noindent{\bf Basic Idea}.
We propose using {\it low-level} and {\it high-level} psychological features to characterize sophistication of malicious emails. Low-level psychological features are visible representations of salient textual or imagery elements in a (malicious) email to increase the likelihood of recipient compliance. These elements could resemble a familiar picture, logo, or keyword that instills a sense of confidence in the recipient.\ignore{ to encourage complacency. by familiar phrases or appearance of expected graphics.} High-level features, by contrast, reflect the overall email framing (i.e., presentation).\ignore{to the recipient as a whole,} For example,\ignore{attempting to exploit a recipient's goal to respond quickly, or} sending a message from a (purported) source of authority (e.g., a supervisor) while amplifying pressure to act under time pressure (e.g., ``We have a problem -  call me immediately''). The low-level features (i.e., elements) and high-level features (i.e., framing) can be seen as the respective counterparts of the attack {\it techniques} and {\it tactics} in MITRE's ATT\&CK framework \cite{mitreattack}. Thus, we refer to the low-level psychological features as {\it Psychological Techniques} (PTechs) and high-level psychological features as {\it Psychological Tactics} (PTacs). 
We use the concepts of PTechs and PTacs to define metrics and quantify sophistication of malicious emails. 

\smallskip

\noindent{\bf PTechs}. A PTech is a concrete (i.e., quantifiable) cue such as a textual or an imagery element that encourages individuals to comply with a social engineering attack. PTechs that have been identified in the literature include  \cite{montanez2022csekc,longtchi2024internet}:

\begin{enumerate}
\item {\sf Urgency}: The use of textual elements (e.g., ``acting now'') to trigger a quick response from the recipient 
\cite{chowdhury2019impact,vishwanath2011people}.
That is, putting a time constraint on recipients to force them to act fast without thoughtfulness.
\item {\sf Visual Deception}: The use of visual elements (e.g., logos) or ``similar'' characters in URL (e.g., replacing 'vv' with 'w', or 'm' with 'rn') to project trust 
\cite{moreno2017fishing,montanez2020human}. The attacker's malice is hidden in plain sight.

\item {\sf Incentive \& Motivator}: The use of textual or graphic elements to indicate a high discount or a freebie such as ``gift cards'' (incentive) or ``help others'', to incentivize or motivate a recipient to take action \cite{beckmann2018motivation,montanez2022csekc}.

\item {\sf Persuasion}: The use of textual elements related to Cialdini's principles of persuasion (e.g., ``C-Suite titles,'' ``last chance,'' or ``expert opinion'') to encourage a recipient to take action \cite{frauenstein2020susceptibility,ferreira2015analysis}. The six principles are: {\sc authority}, which describes power or dominance over someone; {\sc reciprocation} (or {\sc reciprocity}), which describes the tendency to pay back a favor; {\sc liking} (or {\sc similarity}), which describes one's tendency to react positively to those people they have a relationship; {\sc scarcity}, which uses the lack of goods or services to lure victims; {\sc social proof}, which describes one's tendency to imitate others;
{\sc consistency} (or {\sc Commitment}), which describes the extent one is dedicated to a person or something.
 
\item {\sf Quid-Pro-Quo}: The use of textual elements (e.g., ``Pay an upfront fee'') to ask a recipient for a favor in exchange for a bigger reward \cite{stajano2011understanding}. Note that {\sf Quid-Pro-Quo} is different from {\sc reciprocity} because the former 
is about agreeing to pay back before a deed (i.e., similar to palm-greasing or bribery) but the latter 
is about paying back after a deed (i.e., similar to freewill). 

\item {\sf Foot-in-the-Door}: The use of textual  elements (e.g., ``from our last email ...'') to obtain compliance from a recipient via gradually increasing  demands \cite{freedman1966compliance}. Note that it 
is different from politeness because it eventually gains full access by making an individual to gradually accept modest requests.

\item {\sf Trusted Relationship}: The exploitation of an established third-party relationship of trust with the recipient by using textual elements, such as ``John told me about you'' to convince a recipient to take action \cite{allodi2019need}.  

\item {\sf Impersonation}: The use of a false persona to gain the trust of a recipient by using elements, such as ``I'm billionaire Warren Buffet'' \cite{ferreira2015analysis,allodi2019need}. Note that {\sf Impersonation} is different from {\sf Pretexting} because the former takes a fake persona to hide one's real identity or to gain trust but the latter presents a fake narrative/story to gain trust.

\item {\sf Contextualization}: Referencing current event by using textual elements, such as ``the Pandemic'' or ``War in Ukraine'' \cite{MirIEEEISI2020Landsapce,goel2017got,montanez2020human}. The event can be an activity only known to members of a closed community (e.g., work retreat).
    
\item {\sf Pretexting}: Providing a motive to establish contact with a recipient by using textual elements, such as ``I am recruiter for XYZ company'' \cite{alhamar2010ieee,goel2017got}. 

\item {\sf Personalization}: Addressing a recipient using detailed personal information in textual elements, such as ``Dear John'' or ``Your credit card ending in ...'' \cite{hirsh2012personalized,jagatic2007acm}.

\item {\sf Attention Grabbing}: The use of graphical/auditory elements to draw attention to textual elements, such as highlighted text, brightly colored buttons, or extra large fonts \cite{nelms2016usenix,flores2015investigating}. Note that unlike {\sf Visual Deception} where attackers use visuals to instill trust, attackers use visuals in {\sf Attention Grabbing} to draw a recipient's attention to what the attacker wants the recipient to see or do.

\item {\sf Affection Trust}: Developing an effective relationship to extort a recipient by using textual elements, such as ``My child is sick and I have no money to pay for the treatment'' \cite{montanez2022social}.

\item {\sf Decoy Effect}: Making one to believe that something is a good deal (e.g., presenting a user with a lower than the market price for some goods but actually offering a fake one or never delivering when a victim pays upfront) \cite{seitz2016influencing, cui2022does}. 

\item {\sf Priming}: Influencing one's decision through gradual manipulation (e.g., sending them information about cryptocurrency as the next big thing before sending them a fake link to purchase cryptocurrency) \cite{gillath2019attachment}.

\item {\sf Loss Aversion}: Providing something for free, but later charging enormously when a victim becomes attached to the free item (e.g., providing live soccer links, then charging them when they become attached to the free live soccer links) \cite{benias2018hacking}. 
\end{enumerate}

\smallskip

\noindent{\bf PTacs}. 
A PTac is a measure of the overall coherence and quality of the message based on established ideas of framing and relevance
\cite{goel2017got} to influence 
decision-making \cite{montanez2020human}. \ignore{The impact of message framing in cybersecurity and cyber social engineering attacks has been investigated in several studies \cite{anderson2020security,montanez2020human,montanez2022social}.} Each PTac aims to measure an attacker's effort at crafting and framing an email effectively to prompt a recipient's action. \ignore{PTacs reflect the amount of deliberate effort of an attacker in exploiting specific psychological weaknesses.} 
PTacs can be extended to accommodate other psychological factors and framing approaches that may be proposed in future studies. PTacs that have been implicitly, but not explicitly, proposed in the literature include:
\begin{enumerate}
\item {\tt Familiarity}: This refers to how an attacker engenders a positive (and therefore trusting) association with a recipient. Emails of high {\tt familiarity} may impersonate specific people (e.g., co-workers, bosses, family members, close friends)  \cite{montanez2022social,alhamar2010ieee}.

\item {\tt Immediacy}: This refers to the amplification of a time constraint as a mechanism to shortcut a recipient's skepticism or scrutiny for a desired action, for example, by suggesting that promptness, swiftness, or a quick reaction is required \cite{montanez2020human,nelms2016usenix}.

\item {\tt Reward}: This refers to a clear exchange of something (physical or social) valuable for a recipient. Rewards are often presented as a tangible goods (e.g., money) in exchange for action but can be an offer to improve social standing (e.g., authority, prestige) \cite{goel2017got,longtchi2024internet}.

\item {\tt Threat of Loss}: This refers to an appeal to a recipient's desire to maintain their current status, prevent a loss (e.g., opportunity) or injury (e.g., damage, pain), or avoid the risk of having something stolen. Loss has been hypothesized to be more impactful than potential gain (e.g., reward) \cite{goel2017got,stajano2011understanding,ferreira2015analysis}.

\item {\tt Threat to Identity}: This refers to the efforts by an attacker to manipulate a recipient's desire to maintain a positive, socially valuable reputation (e.g., ``Pay your dues or face the consequences'') \cite{stajano2011understanding,montanez2022social}. 

\item {\tt Claim to Legitimate Authority}: This refers to emphasizing a source of legitimate power to obscure or deter increased scrutiny. The attacker may assume a position of technical expertise, a valuable institutional role, or a traditionally respected office \cite{stajano2011understanding,ferreira2015analysis}. Note that it is different from the Principle of Authority 
because a legitimate authority in the former case
does not have to be a human but the authority in the latter case refers to human experts by definition.

\item {\tt Fit \& Form}: This refers to how a message mirrors the expected composition style of an authentic message. An attacker often exploits commonly expected written or visual display format to resonate with the email’s apparent sender and purpose \cite{rajivan2018creative,goel2017got}. Note that unlike {\tt Familiarity} which is about being familiar with the content of an email, {\tt Fit \& Form} is the general expectation on how the email should look and feel when coming from the purported sender.
\end{enumerate}


\section{Framework}\label{Methodology}

The framework is centered around using PTechs and PTacs to measure the sophistication of malicious emails. Intuitively, a malicious email is an email whose objective is hidden from the recipient but beneficial to the sender (i.e., attacker). In other words, a malicious email attempts to coerce a recipient to do something in compliance with the request described in the email, which the recipient would not do if the recipient knew the sender's objective.
The rationale is that each psychological feature represents a different aspect of the attacker's effort. 

Intuitively, PTechs can be seen as {\it quantifiers} of a malicious email content, reflecting the presence of {\it elements} that evidence the attackers' effort; whereas, PTacs can be seen as {\it qualifiers} of malicious email content, reflecting the attackers' overall effort. 
To our knowledge, this is the first work that provides a systematic decomposition of email contents coupled with attacker effort to quantify the sophistication of malicious emails. Moreover, the framework can be extended to accommodate new factors that may be identified by future advancement in psychology research. 


\begin{figure}[htbp!]
\centering 
\includegraphics[width=0.45\textwidth]{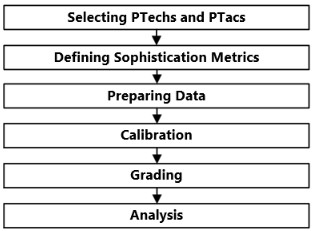}
\caption{Overview of the framework, where the calibration process is iterative. 
}
\label{fig:Framework_flow} 
\end{figure}


The framework consists of six components as shown in Figure \ref{fig:Framework_flow}: (i) selecting PTechs and PTacs for assessment; (ii) defining  metrics to quantify sophistication of malicious emails; (iii) preparing a dataset of malicious emails for expert graders to assess; (iv) calibrating the grading process, including designing grading rules and  training;  (v) grading emails; and (vi) analyzing outcome of the grading process.
These steps may look unfamiliar to computer scientists, who are often given datasets to analyze. 

\ignore{
\begin{figure}[htbp!]
\centering
\includegraphics[width=0.9\textwidth]{images/Flow_diagrams/Framework_flow12.jpg}
\caption{Framework for quantifying sophistication of malicious emails through the lenses of PTechs and PTacs.}

\label{fig:framework}
\end{figure}
}

\subsection{Selecting PTechs and PTacs}\label{ss.select}

We propose selecting the PTechs that (i) are known to be used in malicious emails based on research evidence and (ii) require a one-time interaction to be effective.
This selection criterion is flexible enough to accommodate future understanding and knowledge (e.g., when new PTechs are discovered in the future). Similarly, we propose selecting PTacs that (i) are known to be used in malicious emails based on research evidence, (ii) are independent of one another, and (iii) reflect the holistic effort of an attacker. Suppose, according to the respective selection criteria, some number $\ell$ of PTechs are selected, denoted by $\{\PTech_1,\ldots,\PTech_\ell\}$, and some number $m$ of PTacs are selected, denoted by $\{\PTac_1,\ldots,\PTac_m\}$. This allows us to define sophistication metrics as follows.  %

\subsection{Defining Sophistication Metrics}\label{Metrics}

We propose expressing sophistication of malicious emails through PTech-based and PTac-based analysis and quantifying it as a two-dimensional vector. We consider both the ``ideal world'' where there are no outliers in terms of subjective measurements and the ``real world'' where there are outliers. This is important and inevitable because we need human experts to ``grade'' malicious emails with respect to PTechs and PTacs. This process is analogous to obtaining the ground-truth labels in machine learning, namely that we need ground-truth labels to train models. Moreover, outliers could still be relevant even after having some calibration process that aims to train human graders in agreeing on how to grade. The calibration process is one component of the framework and will be described later.

\smallskip

\noindent{\bf Metrics for Measuring PTechs}. 
To clarify our way of thinking, let us start with the ideal world. Consider a given malicious email and a set of $\ell$ PTechs denoted by $\{\PTech_1,\ldots,\PTech_\ell \}$. For $\PTech_i$ where $1\leq i \leq \ell$, we propose counting the number of elements with respect to $\PTech_i$, leading to an integer grade (or score) $s'_i$. Assuming the $s'_i$ for every $i$ is defined in the same range $[0,\gamma]=\{0,1,\ldots, \gamma\}$. Then, the sophistication of the malicious email through the lens of the $\ell$ PTechs can be defined as,\ignore{the average}
$$
s'=\frac{1}{\ell}\sum_{i=1}^\ell s'_i.
$$

In the real world, the ground-truth $s'_i$ is difficult to obtain.
Thus, we propose approximating it by using a number of $n$ graders (or evaluators) to count the elements concerning $\PTech_i$ while assuring that the graders can count the elements as consistently as possible. For a given malicious email, let $s_{i,j}$ denote the count of elements in the email by grader $j$ with respect to $\PTech_i$, where $1\leq j \leq n$ and $1\leq i \leq \ell$. Then, the sophistication of the email concerning $\PTech_i$ can be defined as,
\begin{equation}
S_i=\frac{1}{n}\sum_{j=1}^n s_{i,j}.
\label{perPTech_calc}
\end{equation}
Given $S_i$ for $1\leq i \leq \ell$, we propose defining the sophistication of the given malicious email with respect to the $\ell$ PTechs, denoted by $S_{\PTech}$, as:
\begin{equation}
S_{\PTech}=\frac{1}{\ell}\sum_{i=1}^{\ell} S_i.
\label{S_PTech_definition}
\end{equation}

Note that Eq.\eqref{def:sophistication} operates by assuming every grade should be considered. In the practice, some grade(s) by some grader(s) may be considered outliers and thus eliminated. This means that Eq.\eqref{def:sophistication},  
should be amended to accommodate the elimination of outliers, while keeping Eq.\eqref{S_PTech_definition} 
intact.
Specifically, when coping with Eq.\eqref{perPTech_calc}, which computes the average grade $S_i$ of $\PTech_i$ by the $n$ graders, we may encounter a subset of graders, denoted by $J'\subset \{1,\ldots,n\}$, being outliers according to some well-established criteria (e.g., the specific one that will be used in our case study, but there could be others), meaning that their grades, namely $s_{i,j}$ for $j\in J'$, 
should be excluded when computing the average grade $S_i$. (Note that $J'=\emptyset$ corresponds to the absence of outliers). As a result, Eq.\eqref{perPTech_calc} becomes for $1\leq i \leq \ell$:
\begin{equation}
S_i=\frac{1}{n-|J'|}\left(\sum_{j=1}^n s_{i,j}-\sum_{j\in J'}s_{i,j}\right).
\label{perPTech_calc-revised}
\end{equation}

\noindent{\bf Metrics for Measuring PTacs}. Similarly, we start with the ideal world. Consider a malicious email and a set of $m$ PTacs denoted by $\{\PTac_1,\ldots,\PTac_m\}$. Since the ground-truth sophistication is difficult to obtain, we propose assessing $\PTac_i$ using a scale $[0,\beta]=\{0,1,\ldots,\beta\}$ for some integer $\beta$,
also by $n$ independent graders, where $p_{i,j}$ denotes the assessment of grader $j$ with respect to $\PTac_i$ for a given malicious email, $1\leq j \leq n$, and $1\leq i \leq m$. The sophistication of the given malicious email with respect to $\PTac_i$ can be defined as,

\ignore{
where by grader $j$ where $1\leq j \leq n$ yielding the following PTac score for each 
{\color{red}
For each PTac scale i, denote by $p_{i,j}$ the assessed attacker effort given  with respect to $PTac_i$. Then, the sophistication of the email with respect to $PTac_i$ can be defined as
{\color{magenta}: Ensure it makes sense -  j is grader and i each of the Ptac}}}
\begin{equation}
P_i=\frac{1}{n}\sum_{j=1}^n p_{i,j}.
\label{perPTac_calc}
\end{equation}
The overall PTac-based sophistication of an email 
 can be defined as:
\begin{equation}
S_{\PTac}=\frac{1}{m}\sum_{i=1}^{m} P_i.
\label{S_PTac_definition}
\end{equation}

Note that Eq.\eqref{perPTac_calc} assumes there are no outlier grades, which need to be eliminated.
Denote by $J''\subset \{1,\ldots,n\}$ the set of outliers,
meaning that their grades, namely $p_{i,j}$ for $j\in J''$, 
should be excluded when computing the average grade $P_i$, where $J''=\emptyset$ corresponds to the absence of outliers.
As a result, Eq.\eqref{perPTac_calc} now becomes for $1\leq i \leq m$:
\begin{equation}
P_i=\frac{1}{n-|J''|}\left(\sum_{j=1}^n p_{i,j}-\sum_{j\in J''}p_{i,j}\right),
\label{perPTac_calc-revised}
\end{equation}
but Eq.\eqref{S_PTac_definition} remains unchanged. 

\smallskip

\noindent{\bf Metrics for Quantifying Sophistication of Malicious Emails}. 
By treating $J'=\emptyset$ as a special case of $J\neq \emptyset$ and treating $J''=\emptyset$ as a special case of $J''\neq \emptyset$, we obtain:
\begin{definition}[sophistication of malicious email] 
\label{def:sophistication} The sophistication of a malicious email is measured as a two-dimensional vector $(S_{PTech},S_{PTac})$, where $S_{PTech}$ is defined in Eq.\eqref{S_PTech_definition} with $S_i$ in Eq.\eqref{S_PTech_definition} being defined in Eq.\eqref{perPTech_calc-revised},
and $S_{PTac}$ is defined in Eq.\eqref{S_PTac_definition} and with $P_i$ in Eq.\eqref{S_PTac_definition} being defined in Eq. \eqref{perPTac_calc-revised}.
\end{definition}

\ignore{
\begin{definition}[sophistication of malicious email] 
\label{def:sophistication} In the absence of outliers, the sophistication of a malicious email is measured as a two-dimensional vector $(S_{PTech},S_{PTac})$, where $S_{PTech}$ is defined in Eq.\eqref{S_PTech_definition} and $S_{PTac}$ is defined in Eq.\eqref{S_PTac_definition}.
In the presence of outliers, 
With the amended Eqs.\eqref{perPTech_calc-revised} and \eqref{perPTac_calc-revised}, Eqs.\eqref{S_PTech_definition} and \eqref{S_PTac_definition}, and thus Definition \ref{def:sophistication}, remain valid.
\end{definition}
}


\subsection{Preparing Data}

Several issues must be addressed when preparing data, including collection and preprocessing.  
First, to ensure dataset quality,
we must assure the emails are malicious as the purpose is to 
quantify their psychological sophistication and show the importance and relevance of considering psychological sophistication in future studies (e.g., different defensive mechanisms may be used to cope with malicious emails with different degrees of psychological sophistication).

Second, we must ensure that the data preparation process does not cause damage to the research environment.  This is important because malicious emails may contain links to executable code or malicious websites that can compromise the experimental environment if accidentally clicked, thus possibly affecting other computers in the network.  Therefore, using a virtual environment to isolate emails from the Internet is imperative when preprocessing malicious emails.

Third, malicious emails in a given dataset may contain broken links or missing images, which are needed for assessing their sophistication because these contents would be presented when a recipient views an email in the real world.  This means that we must reconstruct an email by adding the missing links or images.  For example, if a malicious email is missing the Amazon.com logo, we can reconstruct the email by adding the mission logo.  If we cannot reconstruct an email, the email should not be used in this study. Moreover, the reconstruction process should be sound.
For example, an email dated with year 2006 with a missing PayPal logo must be replaced only with the 2006 PayPal logo.

Fourth, given a set of malicious emails, we must ensure that each email content is rendered similarly, if not exactly the same, on different machines and platforms from a visual point of view.  This is important because an email will be assessed by multiple graders.
This is not trivial to guarantee because graders may use different software platforms, email readers, or web browsers, which might render emails differently on each computer. We ensure this by presenting a screenshots of emails to the graders.


\ignore{\color{red}Fifth, we may need to sanitize emails to prevent the leakage of sensitive information but without removing message content that is relevant for the analysis. {\color{ForestGreen}This means we may redact the sensitive content of an email such as {\color{red}part of the recipient's name or card number,}\footnote{describe clearly what is redacted during the grading process...if nothing is redacted, delete this "Fifth". } leaving parts that clearly show the grader that it is a name or a card number.}} 

\subsection{Calibration}\label{ss.cal}

\begin{figure}[htbp!]
\centering
\includegraphics[width=0.46\textwidth]{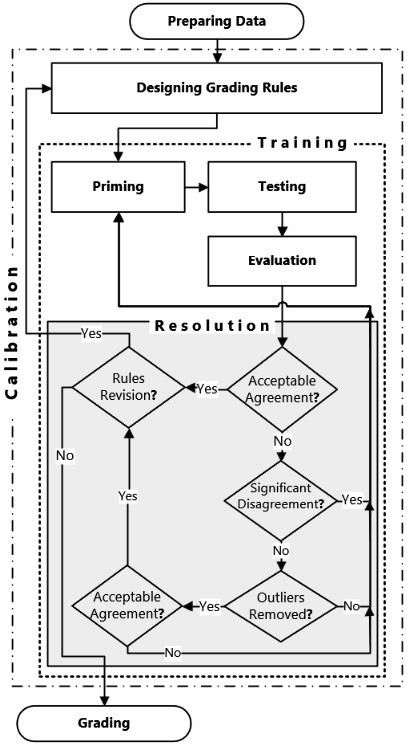}
\caption{\small The Calibration Process includes two sub-processes: {\it Designing Grading Rules} and {\it Training}, where the latter has four steps---priming, testing, evaluation, and resolution.
}
\label{fig:Calibration_Assessment}
\end{figure}

The calibration process aims at mitigating human (including expert) subjectivity in grading sophistication of malicious emails. In a sense, the calibration process is reminiscent of the model training step in machine learning. 
The calibration process has two sub-processes: {\it Designing Grading Rules} and {\it Training}, which are highlighted in Figure \ref{fig:Calibration_Assessment}, including the iteration that may be incurred.
Calibration is important because as mentioned above, we can only approximate the unknown ground-truth metrics by having domain experts 
grade emails with respect to PTechs and PTacs according to  our \textit{grading rules} (in a fashion similar to \cite{wash2020experts,ulqinaku2020real,garcia2020reviewing}).
When a grader manually counts the number of psychological elements exhibited in an email with respect to a PTech and assesses the overall coherence and quality of the email with respect to a PTac, bias and/or subjectivity is inevitable because the interpretation of ``psychological elements'' relies on one's domain expertise. This explains why we need the {\it Designing Grading Rules} sub-process. Even given grading rules, we still need to train graders to eliminate as much bias or subjectivity as possible.

\noindent{\bf The Designing Grading Rules Sub-Process}. The purposes of grading rules is to guide graders in scoring the psychological sophistication, more specifically psychological elements, of emails with respect to PTechs and PTacs. We propose designing grading rules in an iterative fashion as follows.
\begin{enumerate}
\item Initial grading rules are designed by some experts, such as: (i) instructions on recognizing a PTech, how to count its use, and what the metric scale is;
(ii) instructions on scoring the effectiveness of a PTac in an email, and what is the metric scale.
In our case study, we will present specific initial grading rules with respect to each PTech and PTac.

\item The initial rules are used in the {\it Training} sub-process, which (as described below) helps decide whether to revise the current grading rules (e.g., adding new rules or deleting existing rules). If no revision is necessary, the current rules will be used in the subsequent Grading process; otherwise, the current rules will be revised and the revised rules will be used in the subsequent iteration of the Training sub-process (i.e., effectively returning to the previous step).

\item Rule for Removing / Keeping Outlier Grades. When there is a statistically significant variation of agreement among the grades,
we may need to eliminate the outlier grades that are outside of a predetermined acceptable threshold. For this purpose, we need a rule 
to specify: (i) what makes a grade assigned to an email with respect to a specific PTech or PTac  an outlier;
(ii) how to decide whether an outlier will be definitely eliminated or not, which is relevant especially when the number $n$ of graders is small.

\end{enumerate}

Since grading rules may be complicated, a {\it grading aid} may be warranted. Intuitively, a grading aid would display the grading rules to the graders in their course of grading so as to reduce their cognitive load on conducting the task.
This is important because a significant number of PTechs and PTacs are involved, and the relationships between these PTechs and PTacs can be delicate. For example, a grading aid may provide a reference table of key psychological elements corresponding to a PTech or PTac and may further provide examples of each PTech or PTac using example emails.

\smallskip

\noindent{\bf The Training Sub-Process}.
Given the initial or revised grading rules, a group of graders are recruited. The graders learn how to apply these rules to grade emails in the {\it Training} sub-process, which has the following four steps.

\begin{enumerate}
\item {\bf Priming}: At this step, graders learn how to grade emails using the initial or current grading rules, ideally supported by a grading aid as mentioned above. Graders can ask each other questions (e.g., what would count as a psychological ``element'' with respect to a PTech?), resolve disagreements between them, and collectively build a shared understanding of the assessment method. Graders may collectively grade sample emails and discuss them.  Then, the sub-process moves to the {\it Testing} step below.
     
\item {\bf Testing}: At this step, each grader assesses a set of emails independently. The grading environment used in this step should be the same as the one that will be used in the grading process. 
  
\item {\bf Evaluation}: Grades (i.e., scores assigned by graders) resulting from the {\it Testing} step are evaluated for their degree of agreement, which can be measured with some well-established metrics. We advocate using Krippendorff's Alpha (Kalpha) \cite{marzi2024k}, denoted by $\alpha\in [-1,1]$, because it is not only applicable to categorical, ordinal, and interval data, but also robust in the presence of missing data (which can occur when outliers are removed) \cite{zapf2016measuring, hallgren2012computing, krippendorff2011computing}. One advantage of using Kalpha over the {\it standard deviation}, which may be tempting, is that

it is not clear how standard deviation can be used to measure the agreement among graders on many emails with respect to many PTechs and PTacs. Moreover, it is not known how standard deviation can account  for ``agreement by chance'' among the graders,  whereas Kalpha addresses this issue 
leveraging the ratio of the observed disagreement to the expected disagreement 
\cite{krippendorff2011computing}.
The meaning of the Kalpha ($\alpha$) is interpreted as follows \cite{marzi2024k,krippendorff2011computing}: $\alpha = -1$ means absolute disagreement;
$-1 \leq \alpha < 0$ indicates 
a degree of disagreement; $0<\alpha < 0.6$ means an unreliable degree of agreement and a resolution among the graders is required;
$0.6 \leq \alpha < 0.8$ means an acceptable degree of agreement but there may still be a degree of disagreement which may need to be resolved;
$0.8 \leq \alpha < 1$ means a highly reliable degree of agreement (i.e., agreement beyond chance and a solid conclusion can be made from the grades); $\alpha = 1$ means a perfect agreement. These parameter regimes will guide the use of $\alpha$ in the {\it Resolution} step below.

\ignore{below a pre-determined threshold, ....what we do to achieve consensus....} \ignore{\color{red}Alternatively, standard deviation can be used as an effective measure, where standard deviation is defined over ... what.\footnote{clarify using the notations we defined above; e.g., standard deviation of $S_i$'s or what?}}
    
\ignore{\color{red}can not be done in all cases does not all cases can be addressed Several special cases still arise during the outlier removal process. In some cases removing outliers does not improve the standard deviation {\color{red}because ......}. {\color{red}When the grading is split (e.g., two graders gave a 1, and the other two graders gave a 5), outliers are not removed. }}
  
\item {\bf Resolution}: At this step, a resolution is to be made based on the degree of agreement $\alpha$ obtained in the preceding step. 

Corresponding to Figure \ref{fig:Calibration_Assessment}, we consider three cases. (i) If $0< \alpha < 0.6$, meaning there is a significant disagreement 
among the grades, 
the graders return to the {\it Priming} step of the Training sub-process. 
(ii) If $0.6 \leq \alpha < 0.8$, meaning there is a potentially acceptable degree of agreement but also a degree of disagreement, the graders may try to resolve the disagreement. One method is to eliminate outlier grades and see if the resulting $\alpha$ significantly increases. 
If eliminating outliers does not increase $\alpha$ significantly,
the graders return to the {\it Priming} step.  
If eliminating outliers does increase $\alpha$ significantly,
the graders need to decide whether to revise the grading rules or not. If not, they proceed to the {\it Grading} process; otherwise,  they return to the {\it Designing Grading Rules} sub-process. 
(iii) If $\alpha \geq 0.8$, the grades are in highly agreement with each other and proceed to {\it Grading}.

\end{enumerate}

\subsection{Grading}\label{ss:grading}

Having accomplished the calibration process, the graders are ready to grade emails (which are not used in the calibration process). 
In this process, the graders {\it independently} assess each email by assigning PTech grades and PTac grades according to the email, while leveraging the grading aid mentioned above (if present). 
The resulting grades will be used for analysis in the next step.

Though sounding simple, the grading process encounters some issues.
First, we should not assume the graders can memorize the grading rules, which should be presented to the graders in the entire grading process. This can be done, for example, by making the grading process based on online survey, while presenting the grading rules or grading aid to the graders in an easy-to-reference fashion.

Second, we must deal with the issue of graders' fatigue. This is necessary because it is not feasible for each grader to grade a large number of emails with high quality within a consecutive period of time. This can be achieved, for example when designing the online survey mentioned above, by breaking down the grading into smaller sessions each of which only requests a grader to grade a small number of emails.
Third, we must give the graders flexibility in terms of the time frame to finish grading because graders may have different working styles or schedules.
Fourth, we must mitigate the bias that may be introduced in the grading process. For example, even if each grading session has  a small number of emails, it is possible that the quality of grading by a grader may decrease somewhat as the grading proceeds (e.g., fatigue may play a role even within a single session of 100 emails).
To mitigate this factor, we can shuffle the emails in terms of the order they are presented to the graders.

\subsection{Analysis}
This process is to analyze the resulting grades from the {\it Grading} process. To make the analysis process more effective, it would be better to design a set of Research Questions (RQs). Examples of Research Questions (RQs) include the following (while noting that other RQs are possible):
\begin{itemize}
\item RQ1: Which PTechs and PTacs are widely employed?
\item RQ2: What emails are more sophisticated?
\item RQ3: How are PTechs and PTacs correlated? 
\item RQ4: Why emails are inconsistently graded? 
\item RQ5: Does sophistication  evolve with time?
\item RQ6: How do attackers exploit social events?
\end{itemize}

\section{Case Study}\label{Experiment Result and Data Analysis}

Now we present a case study to demonstrate the usefulness of the framework, while using an end-to-end example to show the processes.

\subsection{Selecting PTechs  and PTacs}

Under the guidance of the PTech and PTac selection criteria described in framework (Section \ref{ss.select}), we select the following 8 (out of the 16) PTechs described in Section \ref{sec:preliminaries}:
{\sf Urgency},
{\sf Incentives \& Motivators}, 
{\sf Attention Grabbing}, 
{\sf Personalization},
{\sf Contextualization},
{\sf Persuasion}, 
{\sf Impersonation}, 
and {\sf Visual Deception}. 
In the terminology of the framework, we have $\ell=8$. 
Moreover, we select all the 7 PTacs described in Section \ref{sec:preliminaries}: {\tt  Familiarity}, {\tt Immediacy}, {\tt Reward}, {\tt Threat of Loss}, {\tt Threat to Identity}, {\tt Claim to Legitimate Authority}, and {\tt Fit \& Form}.
In the terminology of the framework, we have $m=7$.

\subsection{Instantiating Sophistication Metrics}\label{ss:metrics-def}
\ignore{\color{magenta}Intersection with 3.2 - Define the measure of PTechs and PTacs}

In order to use the sophistication metric given by Definition \ref{def:sophistication}, which is made general enough to accommodate any metric scale, we need to instantiate the metric scales for the PTechs and PTacs, respectively. 
First, for each PTech, we propose using the scale $[0,7]=\{0,1\ldots,7\}$ or $\gamma=7$, 
for two reasons. (i) When developing the grading aid, we notice that most elements of the PTechs do not go above 7 (with the only exception of {\sf Attention Grabbing}). (ii) By limiting the maximum count of PTechs, we mitigate the damage that can be caused when one PTech dominates the PTech grade. 
Second, for each PTac, we propose using the Rating scale $[0,5]$ (i.e., by setting $\beta=5$) which is \ignore{psychometric} commonly used in psychological studies \cite{jebb2021review}.

\subsection{Preparing Data}\label{Preparing-data}

We focus on three types of malicious emails: (i) Phishing emails, which require a one-time interaction for victimization and include a link or an attachment; (ii) Scam emails, which require multiple interactions for victimization via phone call or email exchange, or request personal information; and (iii) Spam emails, which may not obscure information but often sells a product or service that can be misleading.  

We prepare data according to the framework as follows.
First, to assure that the emails are truly malicious, we use reputable data sources. In total, we collect and use 1,177 emails from two sources: 
\begin{itemize}
\item 1,143 emails from the Anti Phishing Working Group (APWG) repository \cite{apwg2022website}, which collects malicious emails from various sources. These 1,143 emails are selected because they are submitted to APWG by US-CERT (Computer Emergency Response Team), which is a reputable source.

Among these 1,143 emails, 107 will be used in the {\it Calibration} process and 1,036 will be used in the {\it Grading} process.
Since APWG does not tell whether an email is a Phishing, Spam, or Scam, we use the ScamPredictor 
\cite{Scamdocc46:online} to determine an email as Phishing, Spam, or Scam.

\item 34 emails are from a dataset we collected from various sources. We use them because we have investigated them in the past and know what kinds of malicious emails they are. Among the 34, 18 are Phishing emails, 7 are Scam emails, and 9 are Spam emails. These 34 emails are only used in the {\it Calibration} process.

\end{itemize}

We characterize the emails that are respectively used in the {\it Calibration} and {\it Grading} processes as follows:
\begin{itemize}
\item Among the 141 ($=107+34$) emails that are used in the {\it Calibration} process, 73 are used in the {\it Designing Grading Rules} sub-process
and 68 are used in the {\it Training} sub-process.
Among the 73 emails used in the {\it Designing Grading Rules} sub-process, 58.9\% (43) are Phishing emails, 21.92\% (16) Scam emails, and 19.18\% (14) Spam emails. 
The temporal distribution is: 14 emails are from 2022; 8 emails from 2021; 14 emails from 2020; 9 emails from 2019; 2 emails from 2018; 1 email from 2017; 6 emails from 2016; 7 emails from 2015; 1 email each from 2014; 1 email from 2012; 3 emails from 2011; and 7 emails from 2006. 
Among the 68 emails used in the {\it Training} sub-process, 63.24\% (or 43) are Phishing emails, 16.18\% (or 11) are Scam emails, and 20.59\% (or 14) are Spam emails. Their temporal distribution is: 15 emails are from 2023; 42 emails from 2022; and 11 from 2021. This uneven temporal distribution is caused by the fact that we use only the most recent emails 
so that we may piggyback them to help draw insights into the recent trend of malicious email sophistication. 

\item Among the 1,036 emails used in the {\it Grading} process, 64.86\% (672) are Phishing emails, 21.53\% (223) are Scam emails, and 13.61\% (141) are Spam emails. 
The temporal distribution is: 496 emails are from the Decembers of 2006, 2011, 2016, and 2021, where 124 emails are randomly selected for each of these four months;
540 emails are selected September 1, 2021 and August 31, 2022, meaning that 45 emails are randomly selected from each of these 12 months, while noting that the 45 emails selected from December 2021 do not overlap with the 124 emails from the same month mentioned above.
Note that we consider
the Decembers of 2006, 2011, 2016, and 2021 
because December is a holiday season that witnesses the highest malicious emails activities in a year \cite{api2021apwg},
while noting that December 2021 is better represented with 169 ($=124 + 45$) distinct emails. We select September 1, 2021-August 31, 2022 because it was the most recent year with available malicious emails from APWG at the time that we started this study. The random selection is achieved by using the Python NumPy and Pandas libraries.
\end{itemize}

Second, to assure that the malicious emails do not cause damage to the research environment, we use images to present them to the graders. The images are obtained by 
taking screenshots of the emails, while assuring that no link is clicked in the process.

Third, to assure that emails 
are reconstructed without missing elements (e.g., broken images), we make a {\tt .eml} file based on the original raw email header and the body while incorporating the missing element. The emails are sanitized by removing their embedded warnings (e.g., ``Warning!! This email comes from an external source!''), if present. 

Fourth, displaying emails as images also assures that the emails are rendered the same over different computer platforms that may be used by the graders.

\subsection{Calibration}\label{Designing_Grading_Rules}

\subsubsection{The Designing Grading Rules Sub-Process}\label{sss:grading-rules}

This sub-process designs rules to guide the grading of emails with respect to PTechs and PTacs for determining agreement between graders and for resolving disagreements between graders. 
We also describe our grading aid.

\noindent{\bf Rules for Grading with respect to PTechs}.
Grading rules with respect to PTechs are instructions on how to count the psychological elements corresponding to each PTech that is employed in an email. As mentioned above, we set the range for each PTech grade as $[0,7]=\{0,1,\ldots,7\}$, namely $0\leq s_{i,j}\leq 7$ for PTech$_i$.
\begin{itemize}
\item For {\sf Urgency}, count the number of instances that trigger immediate action and force the recipient to act under time pressure (e.g., ``now'', ``immediately'', and ``last chance'').
\item For {\sf Visual Deception}, count the number of visuals that attempt to earn the trust of a recipient, such as logos.
\item For {\sf Incentive \& Motivator}, count the number of instances of external rewards for incentives (e.g., ``job offer'', ``50\% discount''), and count the number of instances of internal rewards for motivators (e.g., ``help me'', ``it's a patriotic thing to do''); 
\item For {\sf Persuasion}, count the number of the following principles \cite{ferreira2015principles} that are used in an email: 
(i) the {\sc Authority} principle, by counting the number of instances of C-Suites titles (e.g., CEO, CFO), instances of providing expert opinion (e.g., from Dr. John Doe), and the number of real logos of known or trusted brands;
(ii) the {\sc reciprocity} principle, by counting the number of instances of demands for repayment of an earlier favor;
(iii) the {\sc liking} principle, by counting the number of instances of portrayal of similarities or common interests (e.g., referencing known people with same objectives); 
(iv) the {\sc scarcity} principle, by counting the number of instances of any demonstration showing a lack or short supply of a goods or service; 
(v) the {\sc social proof} principle, by counting the presence of coercion (e.g.,``As an alumni of...'';
(vi) the {\sc commitment} principle, by counting the number of instances referencing earlier emails and/or the dedication to a cause/activity (e.g., ``We are grateful for you past generosity''). 
Note that as justified above, this metric is upper bounded by 7 even if the actual counts goes beyond 7.

\item For {\sf Impersonation}, count the number of pretence to be another entity (e.g., spoofed email address (domain \& TLD), known personalities). 

\item For {\sf Contextualization}, count the number of presences of contexts that are used to try to establish commonality with recipient, or mentioning relevant current events to engage the recipient (e.g., Covid-19). 

\item For {\sf Personalization}, count the number of direct address of recipient (e.g., recipient's name, email address, telephone number). 

\item For {\sf Attention Grabbing}, count the number of visual or auditory elements that prompt a recipient’s focus (e.g., color button, bold font, upper case letters, highlighted text, bright-color font).
\end{itemize}

\noindent{\bf Rules for Grading with respect to PTacs}. Recall that for each PTac we set $\beta=5$ as we use the rating scale $[0,5]$ to measure the sophistication of an email with respect to a PTac. We propose using the following grading rule:
\begin{itemize}
\item `0' for \textit{not applicable}, namely that a PTac is not employed in the email;  
\item `1'  for \textit{minimal} application, namely that the attacker does consider the PTac but applies it neither clearly nor consistently; 
\item `2' for \textit{light} application of a PTac in an email, namely that the attacker considers the PTac, but with inconsistency, confusion, or lapses/errors in their approach; 
\item `3' for  a \textit{moderate} application of a PTac in an email, namely that the attacker clearly applies the PTac but may still have inconsistencies in their approach; 
\item `4' for a \textit{significant} application of a PTac in an email, namely that the attacker clearly and consistently applies the PTac with minimal errors or lapses;  
\item `5' for an \textit{extraordinary} application of a PTac in an email, namely that the attacker expertly and diligently crafts their email to apply the PTac in a cohesive and thoughtful way.
\end{itemize}

\noindent{\bf Rule for  Measuring the Degree of Agreement between Grades}.
In order to measure the degree of agreement between the graders (i.e., their grades), 
we propose using the Kalpha metric (i.e., $\alpha$) mentioned in the framework to measure the degree of agreement {\it between} the graders with respect to the PTechs or PTacs.
This is reasonable because there could be many emails in question (e.g., 1,036 emails in our case study), meaning that we need to consider the {\it overall} agreement between the graders, and because the $\alpha$ is known to have the capability to capture the overall agreement. Nevertheless, we propose separating the treatment of the PTechs from that of the PTacs, namely that we will measure the degree of agreement among the graders with respect to the PTechs separately from that of the agreement among the graders with respect to the PTacs. This is reasonable because (i) the definition of email sophistication with respect to PTechs is separate from the definition of email sophistication with respect to PTacs and (ii) PTechs are measured using scale $[0,7]$ while PTacs are measured using scale $[0,5]$.
In what follows we focus on the computation of the $\alpha$ with respect to PTechs, because its counterpart with respect to PTacs is treated in the same fashion.

\begin{algorithm}[!htbp]
	\caption{Computing degree of agreement $\alpha$}
 \label{alg:alpha}
INPUT: $\{s_{k,i,j}\}_{k\in [1,h], i\in [1,\ell], j\in [1,n]}$ where $s_{k,i,j} \in V$ is the grade of email $k$ in terms of PTech $i$ as assigned by grader $j$

OUTPUT: $\alpha$

\begin{algorithmic}[1]
			
\STATE{$T \gets (T_{k,i,v}=0)_{1\leq k\leq h, 1\leq i \leq \ell, v \in |V|}$} \COMMENT{{\footnotesize initializing agreement table $T$}}
   \FOR{$k = 1 \text{ to } h$} 
        \FOR{$i = 1 \text{ to } \ell$}
            \FOR{$j = 1 \text{ to } n$}
            \STATE{$v \gets s_{k,i,j}$, where $s_{k,i,j}\in V$} \COMMENT{{\footnotesize for better readability}}
        \STATE{$T_{k,i,v} \gets T_{k,i,v} + 1$}
            \ENDFOR
        \ENDFOR
    \ENDFOR \COMMENT{{\footnotesize $T_{k,i,v} = |\{j: s_{k,i,j}=v\}|$, namely the number of graders that assign grade $v$ to email $k$ in terms of PTech $i$, or (email, PTech)}}
    \STATE{$\bar{T}_{k,i} \gets 0$ for $1\le k\leq h$ and $1\leq i \leq \ell$}\COMMENT{{\footnotesize initialization}}
    \FOR{$k = 1 \text{ to } h$} 
        \FOR{$i = 1 \text{ to } \ell$}
            \FOR{$v \in V$}
            \STATE{$\bar{T}_{k,i} \gets \bar{T}_{k,i}+T_{k,i,v}$} \COMMENT{{\footnotesize{$\bar{T}_{k,i}$ is the number of valid graders with respect to a $(k,i)$ pair of (email, PTech)}}}
            \ENDFOR \COMMENT{{\footnotesize{it is possible $\bar{T}_{k,i}<n$ as some $s_{k,i,j}$ may be invalid (i.e., outlier and not  considered)}}}
        \ENDFOR
    \ENDFOR
    \STATE{$\bar{T}\gets \frac{1}{h\times \ell} \sum_{k=1}^{h} \sum_{i=1}^\ell\bar{T}_{k,i}$} \COMMENT{{\footnotesize average number of valid graders per $(k,i)$ pair of (email, PTech)}}
    \STATE{$\hat{T}_v \gets 0$ for $v\in V$}\COMMENT{{\footnotesize initialization}}
    \FOR{$v \in V$} 
        \FOR{$k = 1 \text{ to } h$}
            \FOR{$i = 1 \text{ to } \ell$}
            \STATE{$\hat{T}_v \gets \hat{T}_v+T_{k,i,v}$} \COMMENT{{\footnotesize{adding the number of valid graders per grade category $v\in V$ for all $(k,i)$ pairs of (email, PTech)}}}
            \ENDFOR
        \ENDFOR
    \ENDFOR
    \STATE{$\hat{T}_v \gets \frac{\hat{T}_v}{h \ell}$ for $v\in V$} \COMMENT{{\footnotesize{average number of valid graders assigning grade $v\in V$ to the $(k,i)$ pairs of (email, PTech)}}}
    \STATE{$\pi_v \gets \frac{\hat{T}_v}{\bar{T}} $}\COMMENT{{\footnotesize{probability a $(k,i)$ pair of (email, PTech) has grade $v$}}}
    \STATE{$T' \gets T \times I_{|V|}$ where $I_{|V|}$ is $|V|\times |V| \times |V|$ identity matrix} \COMMENT{{\footnotesize see \cite{zaiontz17kalpha} for multiplication definition}}
    \FOR{$k = 1 \text{ to } h$} 
        \FOR{$i = 1 \text{ to } \ell$}
            \STATE{$p'_{a|(k,i)} \gets \sum_{v\in V}\frac{T_{k,i,v} \left(T_{k,i,v}'-1\right)}{\bar{T}\left(\bar{T}_{k,i} -1\right)}$} \COMMENT{{\footnotesize average ratio of agreement of graders assigning grade $v$ to a specific $(k,i)$ pair}}
        \ENDFOR
    \ENDFOR    
    \STATE{$p'_a \gets \frac{1}{h \times \ell}\sum_{k=1}^{h}\sum_{i=1}^\ell p'_{a|(k,i)}$} \COMMENT{{\footnotesize average of ratio of agreement among all graders for all $(k,i)$ pairs}}
    \STATE{$p_a \gets \left(1 - \frac{1}{h\ell\bar{T}}\right) p'_a + \frac{1}{h\ell\bar{T}}$} \COMMENT{{\footnotesize normalized average of agreement among all graders for all $(k,i)$ pairs}}
    \STATE{$p_e \gets \sum_{k=1}^{h} \sum_{i=1}^\ell T' \pi_v \pi_v^{{\bf T}}$} \COMMENT{{\footnotesize expected average of agreement when scoring is at random, where $^{{\bf T}}$ means transpose}}

{\RETURN $\alpha \gets \frac{p_a - p_e}{1-p_e}$}
		\end{algorithmic}
\label{alg:computing-alpha}
\end{algorithm}





Recall that in the framework (Section \ref{Methodology}) that for a given email, we use $s_{i,j} \in V$ to denote the grade assigned to the email by grader $j$ with respect to PTech $i$, where where $1\leq i \leq \ell$, $1\leq j \leq n$, and $V= [0,7]=\{0,1,\ldots,7\}$ in our case study. This is sufficient because we only need to state that we need a method for measuring the degree of agreement among the grades of all emails with respect to all PTechs or PTacs (i.e., we separate the treatment of PTechs from that of the PTacs because they deal with different levels of abstraction). In this case study, we need to use a specific method to measure the degree of agreement among all grades of all emails. Since the specific method we employ needs to explicitly represent email identity $k$, where $1\leq k \leq h$ and $h$ is the number of emails in question, we need to extend the notation of $s_{i,j}$ to $s_{k,i,j}$, which is the grade of email $k\in[1,h]$ 
assigned by grader $j$ 
with respect to PTech $i$. Note that the treatment of PTacs is the same, except that the number of PTacs is different from that of PTechs' and the domain $V$ of PTacs is different from that of PTechs'.

Algorithm \ref{alg:computing-alpha} is the specific method for computing the degree of agreement among the grades assigned to all the $h$ emails with respect to the $\ell$ PTechs, denoted by $\alpha$. The algorithm can be trivially adapted to compute the degree of agreement about the grades with respect to the PTacs. The algorithm is based on the method proposed in  \cite{gwet2011krippendorff}, which however does not give an algorithm description; this explains why we deem Algorithm \ref{alg:computing-alpha} as a side product that may help computer scientists understand the method. The algorithm takes the $s_{k,i,j}\in V$'s as input and computes the $\alpha$ as the output.
The basic idea behind the algorithm is the following.

The basic idea of the algorithm is the following.
Lines 1-9 compute an agreement table $T$, or a $h\times \ell \times |V|$ matrix, where $V=\{0,1,\ldots,7\}$ in our case study and $T_{k,i,v}$ is the number of graders that assign the same grade $v \in V$ to email $k$ with respect to PTech $i$. Using $T$, Lines 10-18 compute the average number of {\it valid} graders per $(k,i)$ combination, namely per (email, PTech) combination, denoted by $\bar{T}$, where ``valid'' means that a grader's grade is not excluded; note that a grade can be excluded when it is an outlier, in which case the grade should be replaced with a value that does not belong to $V$ (e.g., -1) so that the algorithm can be executed correctly. 
Line 27 computes the average number of graders that assign grade $v\in V$, denoted by $\hat{T}_v$. 
Line 28 computes the probability that a random grader assigns grade $v\in V$ to a $(k,i)$, namely (email, PTech), pair, denoted by $\pi_v$.
Line 29 computes an adjusted $T$ based on the similarities among the grades \cite{mauboussin21kalpha,marzi2024k}, denoted by $T'$. Lines 30-34 compute the average ratio of agreement for a $(k,i)$ pair, denoted by $p'_{a|(k,i)}$. Line 35 computes the average of ratio of agreement among all the $(k,i)$ pairs, denoted by $p_a'$. Line 36 computes the normalized average ratio of agreement, denoted by $p_a$. Line 37 computes the observed classification probability when grades are randomly assigned, denoted by $p_e$; this explains why $\alpha$ can offset the impact of random grades. Line 38 computes and returns the agreement $\alpha$. 



\smallskip

\noindent{\bf Rule for Identifying and Removing / Keeping Outliers}. In the case the grades are not in acceptable agreement with each other, we need to deal with outliers, which can happen in both the {\it Training} sub-process of the {\it Calibration} process and the {\it Analysis} process. For this purpose, we propose using the following heuristics to identify outliers, which are described with respect to PTechs but can be easily adapted to deal with PTacs. 
Note that there can be many approaches to dealing with outliers. We propose using the following approach because we have small number of graders ($n=4$) and because the approach can reasonably handle the situations we encounter.
We stress that these heuristics are not a standard statistical approach because $n$ is small, and that social science studies do not necessary use standard statistical approach as demonstrated by the use of Kalpha mentioned above.

\begin{table*}
\footnotesize
\centering
\begin{NiceTabular}{clcccccccc}[hvlines] 
     \Block{1-2}{\textbf{}} && \textbf{Grader 1} & \textbf{Grader 2} & \textbf{Grader 3} & \textbf{Grader 4} & \textbf{Min} & \textbf{Max} & \textbf{Spectrum} & \textbf{Std Dev} \\ 
\Block{8-1}{ PTechs} 
& {\sf Urgency} & 	0 & 	0 & 	0 & 	0 & 	0 & 	0 & 	0 & 	0  	 \\ 
& {\sf Visual Deception} & 	0 & 	0 & 	0 & 	0 & 	0 & 	0 & 	0 & 	0  	 \\ 
& {\sf Incentives \& Motivators} & 	1 & 	0 & 	1 & 	2 & 	0 & 	2 & 	2 & 	0.71  	 \\ 
& {\sf Persuasion} & 	0 & 	0 & 	1 & 	1 & 	0 & 	1 & 	1 & 	0.5  	 \\ 
& {\sf Impersonation} & 	2 & 	1 & 	2 & 	1 & 	1 & 	2 & 	1 & 	0.5  	 \\ 
& {\sf Contextualization} & 	0 & 	1 & 	0 & 	1 & 	0 & 	1 & 	1 & 	0.5  	 \\ 
& {\sf Personalization} & 	1 & 	0 & 	0 & 	0 & 	0 & 	1 & 	1 & 	0.43  	 \\ 
& {\sf Attention Grabbing} & 	2 & 	3 & 	3 & 	3 & 	2 & 	3 & 	1 & 	0.43  	 \\ 
\Block{7-1}{ PTacs}
& {\tt Familiarity} & 	1 & 	2 & 	3 & 	2 & 	1 & 	3 & 	2 & 	0.71  	 \\ 
& {\tt Immediacy} & 	0 & 	0 & 	0 & 	0 & 	0 & 	0 & 	0 & 	0  	 \\ 
& \cellcolor{yellow!20}{\tt Reward} & 	\cellcolor{yellow!20}0 & 	\cellcolor{yellow!20}0 & 	\cellcolor{yellow!20}3 & 	\cellcolor{yellow!20}3 & 	\cellcolor{yellow!20}0 & 	\cellcolor{yellow!20}3 & 	\cellcolor{yellow!20}3 & 	\cellcolor{yellow!20}1.5  \\	
& {\tt Threat of Loss} & 	0 & 	0 & 	0 & 	0 & 	0 & 	0 & 	0 & 	0  	 \\ 
& {\tt Threat to Identity} & 	0 & 	0 & 	0 & 	0 & 	0 & 	0 & 	0 & 	0  	 \\ 
& {\tt Claim to legitimate Authority} & 	0 & 	1 & 	0 & 	1 & 	0 & 	1 & 	1 & 	0.5  	 \\ 
& \cellcolor{blue!10}{\tt Fits and Form} & 	\cellcolor{blue!10}2 & 	\cellcolor{blue!10}1 & 	\cellcolor{blue!10}3 & 	\cellcolor{blue!10}0 & 	\cellcolor{blue!10}0 & 	\cellcolor{blue!10}3 & 	\cellcolor{blue!10}3 & 	\cellcolor{blue!10}1.12  	 \\ 
\end{NiceTabular}
\caption{Grades of an email given by $n=4$ graders, where grades with respect to the  {\tt Reward} PTac are a split and grades with respect to the  {\tt Fits \& Form} PTac are a sequence (i.e., 0, 1, 2, 3); none of these grades should be eliminated as discussed in the text.} 
\label{tab:Split-grading-table}
\end{table*}

Recall that for a given email $k$ and PTech $i$, the grades are $s_{i,j,k}$ for $1\leq j \leq n$. Denote by $max_{i,k}$ the highest grade, namely $max_{i,k}=\max(\{s_{i,j,k}\}_{1\leq j \leq n})$; in the case of multiple grades are the highest, we choose an arbitrary one. Denote by $min_{i,k}$ the lowest grade, namely $min_{i,k}=\min(\{s_{i,j,k}\}_{1\leq j \leq n})$; in the case of multiple grades are the lowest, we choose an arbitrary one.
We define the $spectrum$ of the grades, denoted by $\delta_{i,k}$, as $\delta_{i,k}=max_{i,k}-min_{i,k}$. If $\delta_{i,k}< 3$, where `3' is chosen because for the scale $[0,7]$ we have $\lfloor 7/2 \rfloor=3$, we do not need to consider outliers.
If $\delta_{i,k}\geq 3$, we still do not consider outliers under any of the following two scenarios (which stand out because we encountered them in our case study highlighted in Table \ref{tab:Split-grading-table}): 
(i) the grade is a split, meaning that $\lfloor n/2\rfloor$ graders assign one grade and the other $\lceil n/2 \rceil$ graders assign another grade; (ii) 
the grades are a sequence, such as $\{s_{i,j,k}\}=\{0,1, 2, 3\}$. 
If $\delta_{i,k}>3$ and none of the preceding two scenarios occurs, we need to determine which of $max_{i,k}$ and $min_{i,k}$ should be considered an outlier and eliminated. For this purpose, we propose comparing the following two distances: one distance is between the highest grade, $max_{i,k}$, and the second highest grade, denoted by $max_2$, which is the highest grade after removing $max$ from the set $\{s_{i,j,k}\}$ (if there is a tie, we choose an arbitrary one); the other distance is between the lowest grade, $min_{i,k}$, and the second lowest grade, denoted by $min_2$, which is the lowest grade after removing $min$ from the set $\{s_{i,j,k}\}$  (if there is a tie, we choose an arbitrary one). 
Then, we eliminate the one with a longer distance to its nearest neighbor; in the case of a tie, we do not eliminate any of them as outlier. Then, if $max-max_2>min_2-min$, then we treat $max$ as outlier and eliminate it; if $max-max_2<min_2-min$, then we treat $min$ as outlier and eliminate it; if If $max-max_2=min_2-min$, then we do not eliminate any of them.

\begin{figure*}[htbp!]
\centering
\includegraphics[width=\textwidth]{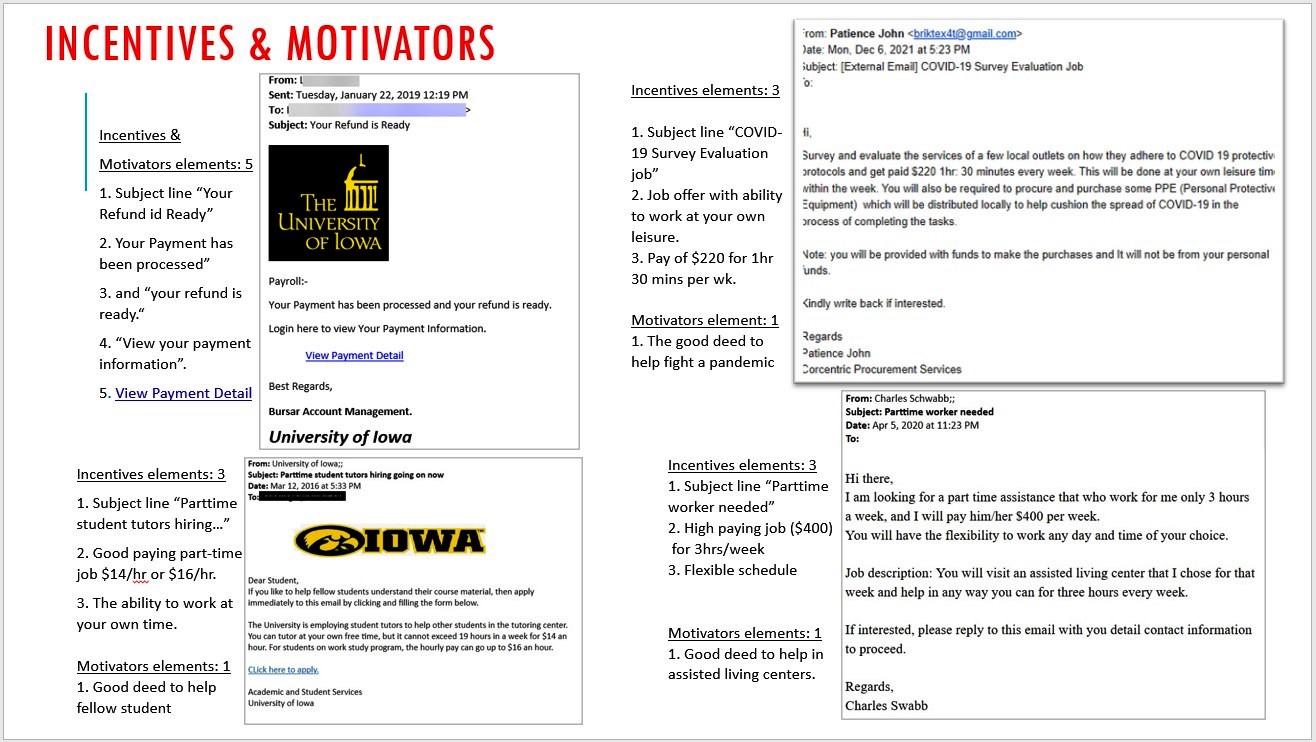}
\caption{{\small Grading Aid Examples. The screenshot of a page in the grading aid document demonstrates how to count {\sf Incentives \& Motivators} cues in emails. For each email, it shows the\ignore{It gives examples of emails with} different counts for cues of {\sf Incentives \& Motivators}, and which cues constitute incentives and which cues constitute motivators. The \textbf{To} field of the emails is redacted for privacy-protection purposes. Note the emails used in the grading aid are not used in the {\it Calibration} or {\it Grading} processes.} 
}
\label{fig:grading_aid} 
\end{figure*}

\noindent{\bf Designing and Developing Grading Aid}.
Guided by the framework, we design and develop a grading aid, which is partly  highlighted in Figure \ref{fig:grading_aid} to show how to count the number of instances of the {\sf Incentives \& Motivators} PTech. 
\begin{itemize}
\item definitions of PTechs and PTacs, and a brief description of how they are employed in an email;

\begin{table*}[!htbp]
\small
\centering
\begin{tabular}{|m{2.4cm} | m{13.8cm}|} 

\hline
\textbf{PTech} & \textbf{Example elements of PTech that may be exhibited in an email} \\
\hline
{\sf Urgency}  & ``call me now" / ``Last chance to save your social life"' / ``This is my last warning" / ``Your PayPal access blocked!" / ``I give your 72 hours to make the payment" \\ \hline

{\sf Visual \newline Deception}  &  PayPal logo, IRS logo / Replacing `womensright.com' with `vvomensright.coom' / Replacing `fbi.gov' with `fbi.gov.net'; Replacing ``microsoft.com" with micros0ft.com \\  \hline

{\sf Incentives \& \newline Motivators} &  ``looking for a part-time assistant,(...) 3 hours a week, (...)\$400 per week" /  ``Your refund notice" /  ``...get paid \$220 1hr 30 minutes every week.\\ \hline

{\sf Persuasion} & {\sc Authority} - ``...Common signs of infection include respiratory symptoms, fever, cough, breathing difficulties..."; {\sc Commitment} - ``We are grateful for you past generosity"; {\sc Liking} ``I'm sure you'll agree with me"; {\sc Reciprocity} - ``We know you appreciate the scholarship you received ..."; {\sc Scarcity} - ``We'll send you a X hat which is limited in supply"; {\sc Social Proof} - ``As an alumni of Calstatela, please consider joining... other alumni who have donated..."   \\ \hline

{\sf Impersonation} & ``Yours sincerely, Warren Buffet" / ``Dr. Eugene Gotcha, World Bank and Trust, 177a Bleecker Street, New York, NY 10012" / Phone/Fax number / CalStateLA Webmail Admin / ``I'm the son of late millionaire president of Nigeria, Umaru Musa Yar’Adua” \\ \hline

{\sf Contextualization}
  & ``following the recent World Standing order over Corona Virus (Covid 19) pandemic..." / ``Your UW.edu account..." / ``The Fed cutting the interest rate to zero'' / ``Government emergency Covid-19 tax relief" / ``...Ken State University retreat..." \\ \hline

{\sf Personalization}  & ``Hi Wendy" / ``Dear Jessica" / ``Important message intended for John Doe" / ``Your home at (home address) / ``Your credit card ending in XXXX\\ \hline

{\sf Attention \newline Grabbing}  & ``{\color{ForestGreen}CLICK HERE}" / ``{\color{blue}Safety Measures.pdf}" / ``{\color{red}Important Covid-19 Updates \& Measures}" / ``\textbf{\underline{Login here to action read}}" / \colorbox{red}{  \color{white}REVIEW NOW  }\\ \hline
\end{tabular}
\caption{Example psychological elements of each PTech 
that can be used as cues exhibited in an email.}
\label{table:PTs_Cues}
\end{table*}

\item examples of malicious emails describing how they are graded with respect to each PTech and PTac; 

\item an instantiated rating scale (i.e., $\beta = 5$) and how to grade the effectiveness of a PTac in an email (i.e., 0 means not applicable, 1 means minimal effort, 2 means light effort, 3 means moderate effort, 4 means significant effort, and 5 means extraordinary effort).

\item a set of real-world emails 
used to show what constitutes a value on the instantiated scale [0, 5] for each PTac as mentioned above. 

\end{itemize}
In addition, the grading aid also includes a reference table of key psychological elements 
associated with a PTech (e.g., the five examples of cues for the {\sf Urgency} PTech and the five examples of cues for the {\sf Attention Grabbing} PTech presented in Table \ref{table:PTs_Cues}, where colors and bold fonts are from real-world examples).

\begin{figure*}[htbp!]
\centering
\includegraphics[width=\textwidth]{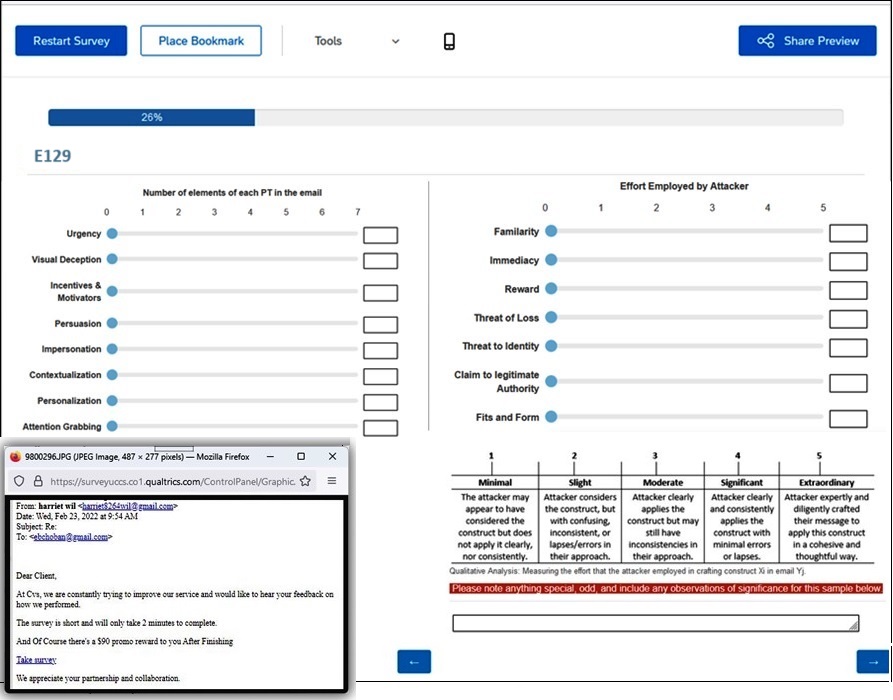}
\caption{\small A modified screenshot showing how a grader sees the survey on the Qualtrics platform. At the top there are options such as restarting the survey. A progress bar indicates the progress with respect to the number of emails for the session (usually 100 emails per session). E129 indicates the ID of the email that is currently being graded. The 8 PTechs are on the left side, and the 7 PTacs are on the right side, both having sliders and text boxes to input counts and grades respectively. The PTacs rating scale is at the bottom right (under the PTacs) to remind graders of what constitutes a grade from 1 to 5. The popup window (a separate window superimposed in this screenshot at the bottom left) portrays the screenshot of the email that is being graded. The popup screen changes to the next random email, but does not change the position of the window. A grader must grade all 8 PTechs and 7 PTacs before proceeding to the next email.} 
\label{fig:Grading-view-Qualtrics} 
\end{figure*}

\subsubsection{The Training Sub-Process}
Four cybersecurity PhD students are recruited to follow the training process described in the framework. 
\begin{enumerate}
\item \textbf{Priming}. At this step, the graders learn the grading rules and the grading aid. The graders are also presented a series of email examples for the graders to assess and discuss as a group. The graders are primed on every item of the grading rules using sample emails. Questions are addressed, and opinions are reconciled. 

\item \textbf{Testing}. At this step, each grader independently assesses a practice set of 52 emails (selected from the 141 emails mentioned above), while noting that $52/1,036\approx 5\%$, which is deemed as an acceptable ratio.
These 52 emails are made available in 
a survey format on the Qualtrics platform as shown in Figure \ref{fig:Grading-view-Qualtrics}. 
The $\alpha$ and standard deviation are computed based on the resulting grades.

\item \textbf{Evaluation}. The graders reconvene to discuss their grades and confirm a shared understanding of all grading rules. 
The $\alpha$ is used to determine the agreement among the graders. Then, the standard deviation is used to see the deviations between the grades, while identifying the PTechs and PTacs grades that exhibit more deviations than others.    

\item \textbf{Resolution}. 
This step addresses and resolves the disagreements between the graders that are identified in the preceding step, as described in the framework.

\end{enumerate}

\subsection{Grading}\label{ss:grading-case-study} 

Our grading process addresses the issues raised in the framework as follows. First, as in the training process, the grading process is conducted using the Qualtrics platform. That is, both the emails and the grading aid are presented as pop-up windows.
Second, to cope with the issue of grader's fatigue, we divide the grading process into 11 self-paced sessions: nine sessions of 100 emails each, one session of 96 emails, and one session of 40 emails (for a total of 1,036).
Third, to give the graders flexibility in the grading process, we ask each grader to conduct each session within 24 hours because we observe that each session of 100 emails may require about 5 hours on average to complete the grading. This gives a grader the flexibility to take breaks between grading sessions.
Fourth, to mitigate the grading bias in terms of the order that emails are presented to graders, we randomize the order of emails within each session (i.e., different graders see emails in different orders). 

\noindent{\bf An End-to-End Example of the Grading Process}.
To help understand how the grading process works, we use one email as as end-to-end example to demonstrate the process. The email is identified as  E129 in our dataset and has an APWG ID\# 116553-5424.24861 in the APWG database. 
In the \textit{Preparing Data} process, we reconstruct the email 
from the raw data collected from APWG,
convert the raw data into a {\tt .eml} file, add the missing logo into the email, and remove warnings message in the email. Then, we open the {\tt E129.eml} file using both Microsoft Outlook and Mozilla Thunderbird to make sure that all the components of the email are correctly displayed in both email clients. Finally, we take a screenshot of E129, and add 
the screenshot to the survey bin in the Qualtrics platform. The screenshot is displayed to the graders in the grading session as shown in Figure \ref{fig:Grading-view-Qualtrics}.
Table \ref{tab:Sample-grading-table} shows the respective grades assigned by the 4 graders to email E129. Results of all the emails is combined and analyze as an aggregate.
\ignore{\color{ForestGreen} The results in Table \ref{tab:Sample-grading-table} is collected together with the results of the grading of the other emails for analysis as a block. That is, the grading results for all the emails for this study are collected and analysed together.} 

\begin{table*}[!htbp]
    \begin{NiceTabular}{clcccccccc}[hvlines] 
     \Block{1-2}{\textbf{PTechs/PTacs}} && \textbf{Grader 1} & \textbf{Grader 2} & \textbf{Grader 3} & \textbf{Grader 4} & \textbf{Min} & \textbf{Max} & \textbf{Range} & \textbf{Std Dev} \\ 
\Block{8-1}{ PTechs} 
   & {\sf Urgency} & 1 & 2 & 0 & 0 & 0 & 2 & 2 & 0.83 \\ 
   & {\sf Visual Deception} & 0 & 0 & 0 & 0 & 0 & 0 & 0 & 0 \\ 
    & {\sf Incentives \& Motivators} & 1 & 1 & 3 & 2 & 1 & 3 & 2 & 0.83 \\ 
   &  {\sf Persuasion} & 1 & 0 & 0	& 2 & 0 & 2 & 2 & 0.83 \\ 
    & {\sf Impersonation} & 2 & 3	& 4 & 4	& 2 & 4 & 2 & 0.83 \\ 
    & {\sf Contextualization} & 0 & 1 & 0 & 0 & 0 & 1 & 1 & 0.43 \\ 
    & {\sf Personalization}	& 0 & 0 & 0 & 0	& 0 & 0 & 0 & 0 \\ 
    & {\sf Attention Grabbing} & 3	& 3 & 2 & 3 & 2 & 3 &  1 & 0.43 \\ 
\Block{7-1}{ PTacs} 
    & \cellcolor{blue!8}{\tt Familiarity} & \cellcolor{blue!8}1 & \cellcolor{blue!8}2 & \cellcolor{blue!8}2 & \cellcolor{blue!8}5 & \cellcolor{blue!8}1 & \cellcolor{blue!8}5 & \cellcolor{blue!8}4 & \cellcolor{blue!8}1.50 \\ 
    & {\tt Immediacy} & 1	& 2 & 0	& 0 & 0 & 2 & 2 & 0.83 \\ 
    & {\tt Reward}	& 2 & 2 & 2 & 2	& 2 & 2 & 0 & 0 \\ 
    & {\tt Threat of Loss}	& 0 & 0	& 0 & 0	& 0 & 0 & 0 & 0 \\ 
    & {\tt Threat to Identity} & 0	& 0 & 0	& 0 & 0 & 0 &0 & 0 \\ 
    & {\tt Claim to legitimate Authority} & 0 & 0 & 2 & 0 & 0 & 2 & 2 & 0.87 \\ 
    & {\tt Fits and Form} & 2 & 0 & 2 & 1 & 0 & 2 & 2 & 0.83 \\ 
\end{NiceTabular}  \caption{Grades of an example email given by $n=4$ graders.
The {\tt Familiarity} PTac 
has a spectrum $\delta_{i,k}=4$, with standard deviation $\sigma=1.5$. 
In this case, grade 5 is an outlier and eliminated according to the outlier removal rule, leading to a smaller spectrum $\delta_{i,k}=1$ and smaller $\sigma= 0.47$.}
\label{tab:Sample-grading-table}
\end{table*}

\smallskip

\ignore{\noindent{\bf Interrater Reliability}.\footnote{this does not appear to be descibed in the framework ... again, we cannot introduce anything here that is not described in the framework...revise}} 

\subsection{Analysis}

The 1,036 emails lead to 62,160 grades as each email is graded with respect to 8 PTechs and 7 PTacs and we have 4 graders.
Out of the 62,160 grades, there are 2,384 (i.e., 15.34\%) outliers which are removed according to the outlier rule mentioned above 
corresponding to a PTech or PTac. 
Among the 2,384 outliers, 
1,001 (3.02\%) are out of the 33,152 PTech grades (i.e., 8 PTechs $\times$ 1,036 emails $\times$ 4 graders) 
and 1,383 (4.77\%) are out of the 29,008 PTacs grades (i.e., 7 PTacs $\times$ 1,036 emails $\times$ 4 graders). 
We observe that 490 of the 1,036 emails had at least one outlier grade removed,
and that split grades occur to 11 out of the 8,288 PTech grades (i.e., 8 PTechs $\times$ 1,036 emails) and 39 out of 7,252 PTac grades (i.e., 7 PTacs $\times$ 1,036 emails). 
In total, 1,001 outlier PTech grades are eliminated and 1,383 outlier PTac grades are eliminated. Before eliminating these outlier grades, we have $\alpha=0.712$ for PTech grades and $\alpha=0.605$ for PTac grades. According to the framework, we proceed to eliminate outliers, leading to 1,001 outlier PTech grades and 1,383 outlier PTac grades. After eliminating these outliers, we observe the $\alpha$ increased, with 
$\alpha = 0.822$ for PTech grades and $\alpha=0.768$ for PTac grades, 
which are used because they represent very reliable agreement as described in the framework.
The discrepancy between the agreement of the PTech grades (i.e., $\alpha = 0.822$) and the agreement of the PTac grades (i.e., $\alpha=0.768$) can be attributed to the fact that PTac grades are more subjective,
as shown by the fact that we eliminate more PTac outliers than PTech outliers.

\begin{figure}[htbp!]
\centering
\includegraphics[width=0.47\textwidth]{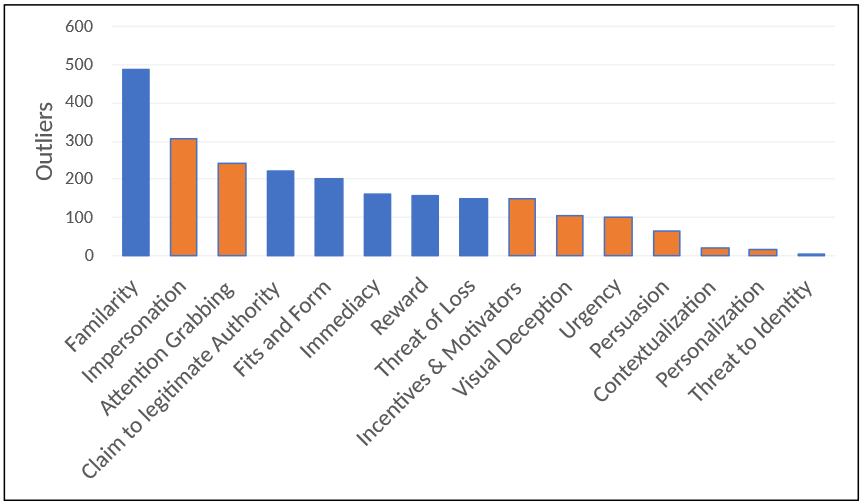}
\caption{\small The total number of outliers for each PTech (orange color) and PTac (blue color). 
} 
\label{fig:Outliers-per-construct} 
\end{figure}

Figure \ref{fig:Outliers-per-construct} plots the 
number of outliers with respect to each PTech and PTac. 
We observe the most outliers are incurred by the {\tt Familiarity} PTac, the {\sf Impersonation} PTech, and the {\sf Attention Grabbing} PTech (in decreasing order), and that the least outliers are incurred by the {\tt Threat to Identity} PTac, the {\sf Personalization} PTech, and the {\sf Contextualisation} PTech (in increasing order).
The implication is important: We need more training and research on the PTechs and PTacs so that graders can identify them more consistently. 
For graders, more consistent grading of malicious emails can better serve as the ground-truth in labelling the sophistication of malicious emails. For machine learning researchers, more training on PTechs and PTacs would allow them to more effectively define features to train machine learning models.
\bigskip
\begin{insight}
{\it To obtain ground-truth PTech and PTac sophistication, we need to provide further training and conduct further research so that graders can adequately approximate the ground-truth sophistication of malicious emails.}
\end{insight}

\subsubsection{RQ 1: Which PTechs and PTacs Are Widely Used?}\label{ss:RQ1}
 
\begin{figure}[!htbp]
\centering
\begin{subfigure}{0.48\textwidth}
  \centering
\includegraphics[width=\linewidth]{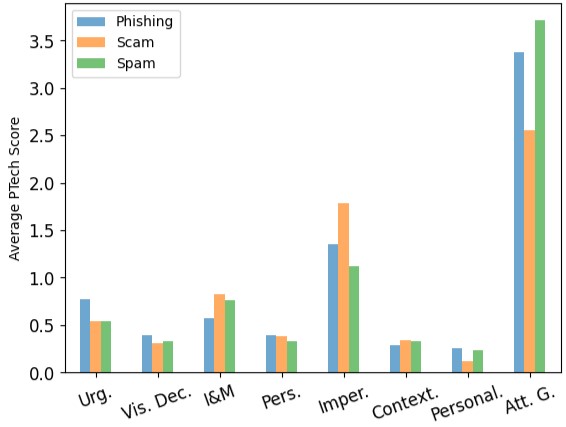}
  \caption{Average PTech grade}
  \label{fig:ptec_count}
\end{subfigure}
\begin{subfigure}{0.48\textwidth}
  \centering
\includegraphics[width=\linewidth]{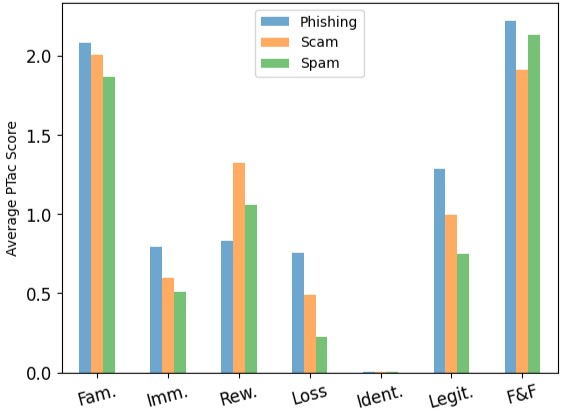}
  \caption{Average PTac grade}
  \label{fig:ptac_count}
\end{subfigure}
\caption{Average PTech score (grade) and average PTac score (grade), where average is among the valid graders of the 1,036 emails.}
\label{fig:occurence}
\end{figure}

To identify the PTechs and PTacs that are most widely employed in each type of malicious emails, Figure \ref{fig:ptec_count} plots the average PTech grade, namely the average of the $S_i$'s defined in Eq.\eqref{perPTech_calc-revised} over the 1,036 emails, where for each PTech the score is the average of the {\it valid} grades (i.e., the grades remaining after eliminating outliers) over the number of {\it valid} graders whose grades are not eliminated as outliers.
Figure \ref{fig:ptac_count} plots the average of the PTac grades, namely the average of the $P_i$'s defined in Eq.\eqref{perPTac_calc-revised} over the 1,036 emails, where the score for each PTac is the average of the valid grades over the number valid graders.

\begin{figure}[htbp!]
\centering
\includegraphics[width=0.48\textwidth]{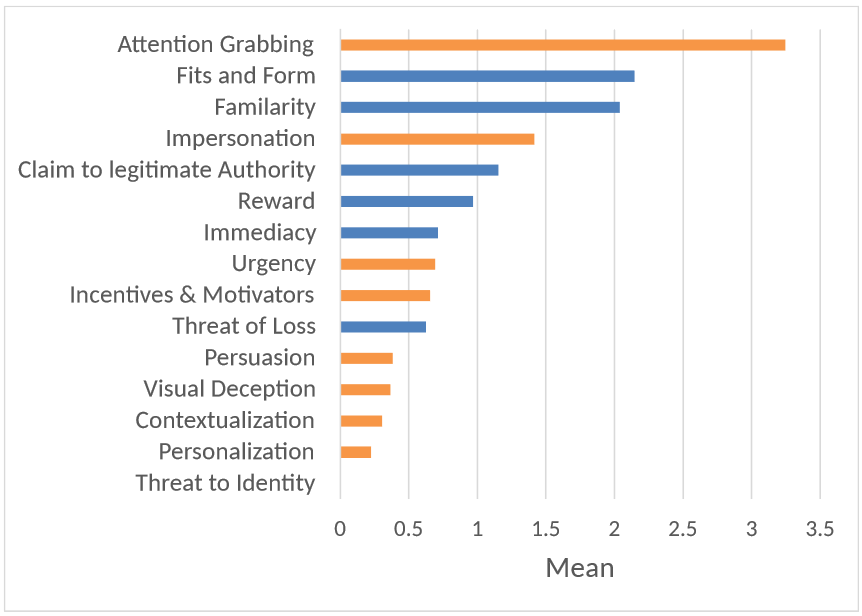}
\caption{{\small Mean value of all 15 constructs showing {\sf Attention Grabbing} (PTech) with the highest mean value across all email types, followed by {\tt Fit \& Form} and {\tt Familiarity} (PTacs).} 
}
\label{fig:PTech_PTac_mean} 
\end{figure}

To see the overall use of PTechs and PTacs,
Figure  \ref{fig:PTech_PTac_mean} plots the mean number of use of the 15 constructs across the three types of malicious emails. We make the following observations. First, {\sf Attention Grabbing} is the most widely employed PTech in all the three types of malicious emails, and is more widely employed in Spam emails than in Phishing and Scam emails. The {\sf Impersonation} PTech is the second most employed PTech and distantly follows the {\sf Attention Grabbing} PTech, and is more employed in Scam emails than in Phishing and Spam emails. 
The third most employed PTech does not follow the pattern of the first two most employed PTechs across all types of malicious email. While the {\sf Urgency} PTech is the third most employed PTech in Phishing emails, the {\sf Incentives \& Motivators} PTech is the third most employed PTech in Scam and Spam emails. Similarly, the {\sf Incentives \& Motivators} PTech is the fourth most employed PTech in Phishing emails, but the {\sf Urgency} PTech is the fourth most employed PTech in Scam and Spam emails. The least employed PTech is the {\sf Personalization} PTech across all three types of malicious emails, perhaps because it is easier for attackers to send out a single generalized email to a large number of individuals than personalizing an email to a single individual. 
It is interesting to note that the {\sf Persuasion} PTech, which is perhaps the most studied PTech in academic literature, is the fifth most widely employed PTech across the three types of malicious emails. 
This may explain why existing defenses are not as effective as desired because of the focus on the {\sf Persuasion} PTech, which is not among the most widely employed PTechs.

Second, the {\tt Fit \& Form} PTac is the most widely employed PTac in Phishing and Spam emails, but the second most employed PTac for Scam emails. It is closely followed by the {\tt Familiarity} PTac, which is the most widely employed PTac in Scam emails. The {\tt Reward} PTac is the third most employed PTac, but is most employed in Scam emails than in Spam and Phishing emails. {\tt Threat to Identity} is the least employed PTac with an extraordinary low occurrence in all three types of malicious emails. 
\bigskip
\begin{insight}
{\it Existing studies might have focused on coping with the less proliferated PTechs (e.g., {\sf Persuasion}) and PTacs (e.g., {\tt Reward}), rather than the most proliferated PTechs (e.g., {\sf Attention Grabbing} and {\sf Impersonation}) and PTacs (e.g., {\tt Fit \& Form} and {\tt Familiarity}).}
\end{insight}


\ignore{

\begin{insight}\label{insight:most_employed_constructs} 
{\it Attention-grabbing} is the most widely employed PTech, while {\tt Fit \& Form} and {\it Familiarity} are the two most widely employed PTac.
\end{insight}

Due to the fact that {\it Familiarity} is so widely employed as indicated in Insight \ref{insight:most_employed_constructs}, and also the fact that {\it Familiarity} is the construct with the most outliers as shown in Figure \ref{fig:Outliers-per-construct}, we draw insight \ref{insight:familiarity_crutiality}.

\begin{insight}\label{insight:familiarity_crutiality}
    {\tt Familiarity} is crucial in determining the sophistication of malicious emails, since it is an important factor that portrays individual difference, which is a characteristic that determines how individuals perceive malicious emails. 
\end{insight}

}

\subsubsection{RQ 2: What emails are more sophisticated?}
To understand how PTechs and PTacs may differ among the three types of malicious emails, we use the z-score method \cite{nield2022essential} to normalize the PTech grades to a scale comparable to the PTac grades. The z-score  replaces a value by subtracting the mean value from it; dividing the result by the standard deviation. The mean values of the normalized PTech grades (i.e., 0.0425 for Phishing emails,  0.0258 for Scam emails, and -0.1443 for Spam emails) are more diverse than the mean values of the PTac grades, which concentrate in a small interval (i.e.,  1.1418 for Phishing, 0.9351 for Scam, and 1.0472 for Spam). This difference suggests that PTech and PTac capture different aspects of malicious emails (i.e., both are important).

\begin{figure}[!htbp]
\centering
\begin{subfigure}[t]{0.235\textwidth}
  \centering
\includegraphics[width=0.95\textwidth]{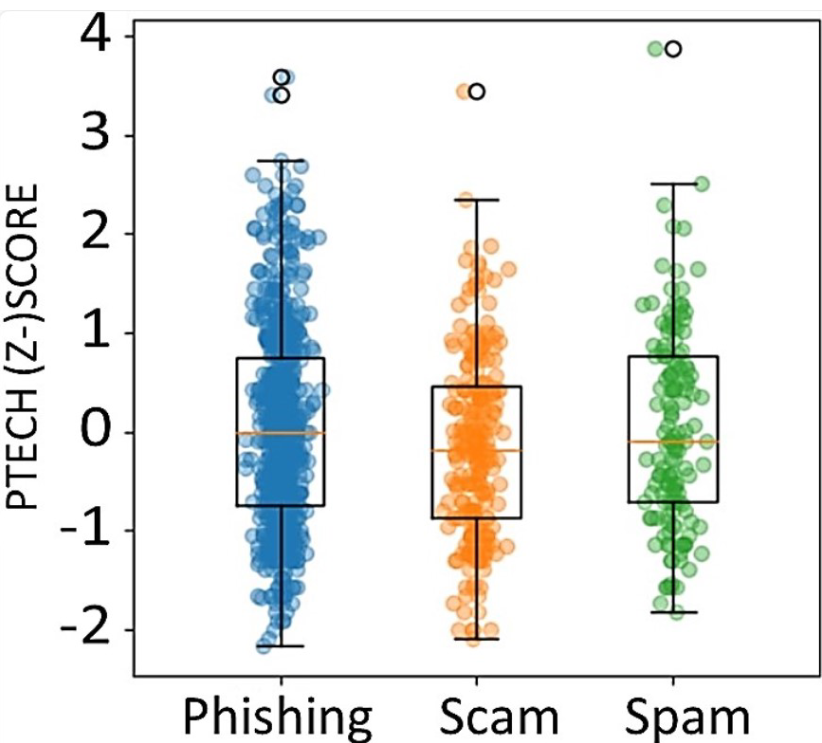}
  \caption{PTech grades.}
\label{fig:ptec_zscore}
\end{subfigure}
\begin{subfigure}[t]{0.235\textwidth}
  \centering
\includegraphics[width=0.95\textwidth]{images/PTech_zscore2.png}
  \caption{PTac grades.}  \label{fig:ptac_zscores}
\end{subfigure}
\caption{{\small Boxplots of the normalized PTech grades and the original PTac grades. 
}}
\label{fig:scores}
\end{figure}

Figure \ref{fig:ptec_zscore} further shows an almost equal z-score
between Phishing and Spams above Scams, and Figure \ref{fig:ptac_zscores} 
shows that Phishing emails exhibit a higher z-score that Scam and Spam emails.
Cognisant of the fact that the sophistication of a malicious email is a two-dimensional vector $(S_{PTech},S_{PTac})$, where $S_{PTac}$ measures the overall coherence and quality of the message, a high PTac z-score for Phishing coupled with a high PTech score makes Phishing emails more sophisticated than the other two kinds of malicious emails. This leads to:

\bigskip
\begin{insight}
{\it Phishing emails are psychologically more sophisticated than Spam and Scam emails.}
\label{ins:Phishing-more-sophisticated}
\end{insight}

\subsubsection{RQ 3: 
How are PTechs and PTacs Correlated?}

\begin{figure*}[ht!]
\centering
\includegraphics[width=\textwidth]{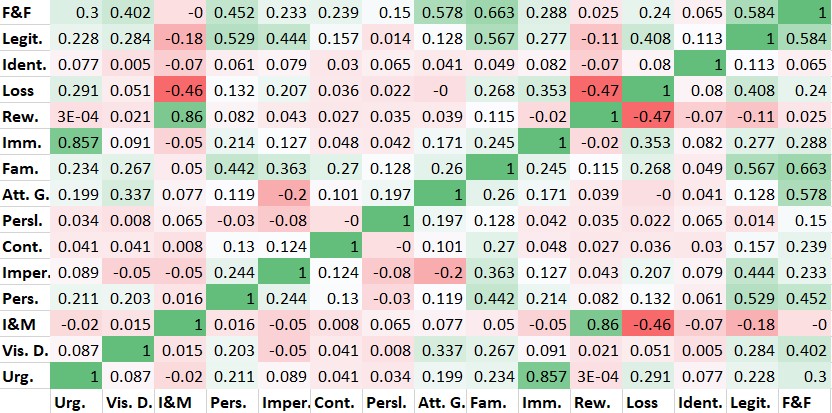}
\caption{\small Correlations between the PTechs and PTacs presented in both values and colors. The darker the color (i.e., green for positive and red for negative) the stronger the correlation. The diagonal green line indicates each PTech and PTac has a 100\% correlation with itself. The very small negative correlation values are seen as -0. The abbreviations are {\sf Urgency} (Urg.), {\sf Visual Deception} (Vis. D.), {\sf Incentives \& Motivators} (I\&M), {\sf Persuasion} (Pers.), {\sf Impersonation} (Imper.), {\sf Contextualization} (Cont.), {\sf Personalization} (Persl.), {\sf Attention Grabbing} (Att. G.), {\tt Familiarity} (Fam.), {\tt Immediacy} (Imm.), {\tt Reward} (Rew.), {\tt Threat of Loss} (Loss), {\tt Threat to Identity} (Ident.), {\tt Claim to legitimate Authority} (Legit.), {\tt Fits and Form} (F\&F). 
} 
\label{fig:Correlation_heatmap} 
\end{figure*}

To answer this RQ, Figure \ref{fig:Correlation_heatmap} summarizes the Pearson  coefficient \cite{schober2018correlation}, which measures the correlation between the PTech grades, between the PTac grades, and between the PTech grades and the PTac grades.

\noindent{\textbf{PTech-PTech Correlations}}. The correlation between the PTechs is weak, with about an equal distribution of negative and positive correlation coefficients falling in between -0.2 and 0.3. The highest positive correlation among the PTechs is between {\sf Attention Grabbing} and {\sf Visual Deception} with a coefficient of 0.337, perhaps because {\sf Visual Deception} requires some elements of {\sf Attention Grabbing} to be effective (i.e., attackers may use visual elements to draw recipient's attention, such as displaying Dropbox logo with a button beneath that reads \textbf{Click here to download}, while the URL actually redirects to a malicious website rather than the Dropbox intended by the recipient). 
The second and third highest positive correlations are between {\sf Persuasion} and {\sf Impersonation} (0.244) and between {\sf Persuasion} and {\sf Visual Deception} (0.202), respectively. 
There are also negative correlations between the PTechs, with the highest negative correlation being between {\sf Impersonation} and {\sf Attention Grabbing}, with a -0.202 coefficient of correlation. This may be because {\sf Impersonation} usually mimics an entity that does not use the elements of {\sf Attention Grabbing} in their emails. 
The remaining negative correlations are too weak. 
\bigskip
\begin{insight}
{\it There is no strong correlation between the PTechs, suggesting that they are relatively independent of each other.}
\end{insight}
\bigskip
\smallskip

\noindent{\textbf{PTac-PTac Correlations}. The correlations between PTacs are generally stronger than the correlations between the PTech. The distribution of the correlations is between -0.466 and 0.663. The highest positive correlation is between {\tt Familiarity} and {\tt Fit \& Form} with a correlation coefficient of 0.663. This correlation may be due to the fact that attackers want the recipient of the email to lack suspicion about the email. Therefore, they make  an email to fit the recipient's expectation of how such an email should look and feel. The second highest positive correlation among PTacs is between {\tt Fit \& Form} and {\tt Claim to Legitimate Authority}, with a coefficient of 0.584. This may be due to the fact that attackers need to frame the email to fit the expectation of the authority that the attacker claims in order to curtail the suspicion of the email recipient. The third highest positive correlation among PTacs is between {\tt Familiarity} and {\tt Claim to Legitimate Authority}, with a coefficient of 0.567. This can be explained by the fact that claiming an authority is worthless if the recipient is not familiar with that authority. Therefore, to succeed in their attacks, attackers tend to claim authorities that are familiar to the email recipients.

The highest negative correlation among PTacs is between {\tt Reward} and {\tt Threat of Loss}, with a correlation coefficient of -0.466. This may be due to the fact that using threats to present a reward has the opposite effect; the greater the threat, the lesser the intended reward. It may also be due to the fact that presenting a reward through threat is counter-intuitive. Therefore, when one of these two PTacs is employed in an email, the other is absent, especially since a reward has to be softly presented as a bait rather than a threat.  
\begin{insight}
{\it {\tt Claim to Legitimate Authority} is the most correlated PTac with other PTacs, probably because attackers tend to claim a legitimate authority that is familiar to the email recipient together with other PTacs.}
\end{insight}
\bigskip

\ignore{\begin{insight}
{\ignore{In a malicious emails PTacs provide a complementary, but different assessment compared to PTechs ,}\footnote{this contradicts our theory, which says that PTechs and PTacs reflects different/complementary things} and the overall sophistication since the number of PTechs employed in an email is not a true determinant of the sophistication.\footnote{does this mean we should only use PTacs to measure sophistication? if so, admit that our hypothesis/theory is invalidated/rejected}} 
\end{insight} }

\noindent{\textbf{PTech-PTac Correlations}}.
The highest positive correlation between PTechs and PTacs is between the {\sf Incentive \& Motivator} PTech and the {\tt Reward} PTac,  with a correlation coefficient of 0.860. This is very closely followed by the correlation between the {\sf Urgency} PTech and the {\tt Immediacy} PTac, with a correlation coefficient of 0.857. The former correlation may be due to the fact that attackers usually present a tangible goods such as money (i.e., a reward for doing something) or ``free stuff,'' or ``help others in need''  in order to incentivize or motivate a recipient to take action. 
The latter may be due to the fact that both the {\sf Urgency} PTech and the {\tt Immediacy} PTac involve time, and whenever the {\sf Urgency} PTech is employed in an email, it triggers the employment of the {\tt Immediacy} PTac. The third highest correlation is between the {\sf Attention Grabbing} PTech and the {\tt Fit \& Form} PTac, with a correlation coefficient of 0.578.
There is a also strong positive correlations between the {\sf Persuasion} PTech and the {\tt Claim to Legitimate Authority} PTac (0.529), and between the {\sf Persuasion} PTech and the {\tt Fit \& Form} PTac (0.4525). 
The highest correlation is between the {\sf Incentive \& Motivator} PTech and the {\tt Reward} PTac, closely followed by correlation between the {\sf Urgency} PTech and the {\tt Immediacy} PTac.

The highest negative correlation is between the {\sf Incentive \& Motivator} PTech and the {\tt Threat of Loss} PTac, with a correlation coefficient of -0.464. This may be due to the fact that an incentive may not achieve its goal if it is presented in an email using threats. This may also be due to  human nature, that a higher degree of threat triggers a recipient's higher degree of analytic reasoning and thus a higher degree of detecting that the the email is malicious. Therefore, attackers avoid using threat in an email if they want to motivate recipients to respond positively to the attackers' requests. The second highest negative correlations is between the {\sf Incentive \& Motivator} PTech and the {\tt Claim to Legitimate Authority} PTac with a correlation coefficient of -0.183, which is weak correlation. This may be because such emails mostly portray the power of authority rather than the skill to motivate. The PTech-PTac correlation shows that less sophisticated emails from the PTech perspective are also less sophisticated from the PTac perspective.

\ignore{
To further explore the relationship between PTech-incurred sophistication and PTac-incurred sophistication, we look at the size of the intersection set of the two sets of emails corresponding to PTech and PTac, respectively. At Q1, the size of the intersection set is: 17.27\% for Phishing, 22.22\% for Scam, and 51.85\% for Spam. At Q3, the size of the intersection set is: 15.45\% for Phishing, 3.17\% for Scam, and 7.40\% for Spam. 
This leads to:}

\bigskip
\begin{insight}  
\label{insight:less-sophisticated-perspective}
{\it There is a strong correlation between the employment of PTechs and PTacs in malicious emails.}  
\end{insight}

\subsubsection{RQ 4: Why emails are inconsistently graded?
}
To answer this RQ we reexamine two outcomes from grading the emails: (i) emails with standard deviations ($\sigma \geq 2$);
and (ii) emails with split grades.

\noindent{\textbf{Emails with high standard deviations ($\sigma \geq 2$)}}. 
There are 159 emails whose PTech or PTac grades have a $\sigma \geq 2$. A further analysis of these emails shows that they are highly charged with the {\sf Impersonation} PTech, with 54.72\% (87 out of 159) occurrences.
The {\sf Impersonation} PTech is to make a recipient think the attacker is someone they know. Therefore, this PTech can skew an email to be high or low depending on the grader's familiarity with the persona that the attacker assumes in the email. This large standard deviation can make an email to be graded high or low depending on: (i) the persona that the attacker assumes; and (ii) the familiarity of the grader with the persona. This can result in an email graded high by some graders and low by other graders. Note also that the persona the attacker assumes can also be an entity such as a known company or any such group. 



\begin{figure}[htbp!]
\centering
\begin{subfigure}[t]{0.23\textwidth}
  \centering
\includegraphics[width=1.01\textwidth]{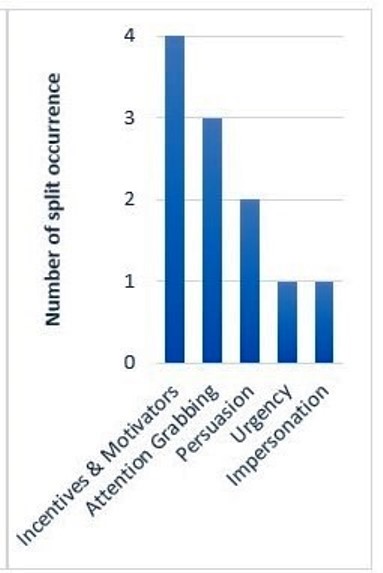}
  \caption{Split PTech grades.}
  \label{fig:Split-occurrence-PTech}
\end{subfigure}
\begin{subfigure}[t]{0.23\textwidth}
  \centering
\includegraphics[width=0.98\textwidth]{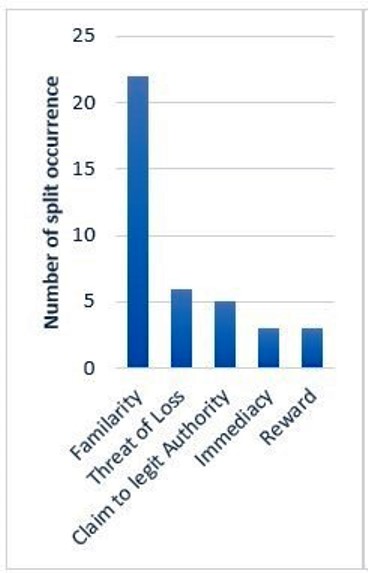}
  \caption{Split PTac grades.}
  \label{fig:Split-occurrence-PTacs}
\end{subfigure}
\caption{{\small Split grade occurrences for PTechs and PTacs, where the $x$-axis shows the PTech or PTac that has at least one split grade. 
}} 
\label{fig:Split-occurrence} 
\end{figure}

\noindent{\textbf{Emails with split grades}}. We reexamine the emails with split grades. Out of 50 split grades occurring to 49 emails (i.e., one email had two split grades), 11 split grades occur to PTechs and 39 occur to PTacs. Figure \ref{fig:Split-occurrence} plots the number of split occurrences in PTech and PTac grades. We observe that 5 (out of the 8) PTechs, as indicated in the $x$-axis, has at least one split grade, and the {\sf Incentive \& Motivator} PTech has 
4 (out of the 11) split grades. We observe that 5 (out of the 7) PTacs, as shown in the $x$-axis, has at least one split grade, among which {\tt Familiarity} has 
22 (out of the 39) split grades and the {\tt Threat of Loss} PTac has 
6 (out of the 39) split grades. 

\begin{table}[htbp!]
\centering
    \begin{NiceTabular}{clcccc}[hvlines] 
     \Block{1-2}{\textbf{PTechs/PTacs}} && \textbf{Phishing} &	\textbf{Scam} &	\textbf{Spam} &	\cellcolor{blue!5}\textbf{Total} \\
\Block{8-1}{\rotate PTechs} 
&	{\sf Urgency} &	1&	&	&	\cellcolor{blue!5}1	\\
&	{\sf Vis. Dec.} &	&	&	&		\cellcolor{blue!5} \\
&	I \& M &	4&	&	&	\cellcolor{blue!5}4	\\
&	{\sf Persuasion} &	1&	&	1&	\cellcolor{blue!5}2	\\
&	{\sf Impersona} &	&	&	1&	\cellcolor{blue!5}1	\\
&	{\sf Context.}  &	&	&	&	\cellcolor{blue!5} \\ 
&	{\sf Personal.} &	&	&	&	\cellcolor{blue!5} \\
&	{\sf Att. Grab.} &	2&	1&	&	\cellcolor{blue!5}3	\\
\Block{7-1}{\rotate PTacs}
&	{\tt Familiarity} &	15&	4&	3&	\cellcolor{blue!5}22	\\
&	{\tt Immediacy} &	&	2&	1&	\cellcolor{blue!5}3	\\
&	{\tt Reward} &	&	&	3&	\cellcolor{blue!5}3	\\
&	{\tt Loss} &	6&	&	&	\cellcolor{blue!5}6	\\
&	{\tt Identity} &	&	&	&	\cellcolor{blue!5} \\
&	{\tt Legit Auth} &	2&	1&	2&	\cellcolor{blue!5}5	\\
&	{\tt F\& F} &	&	&	&	\cellcolor{blue!5} \cellcolor{blue!5} \\
\Block{1-2}{ \cellcolor{orange!10}Total by \\ \cellcolor{orange!10} Email type (\%)}
&	&	\Block{-1}{\cellcolor{orange!10}31 \\ \cellcolor{orange!10} (62\%)} &	\Block{-1}{\cellcolor{orange!10}8 \\ \cellcolor{orange!10}(16\%)} &	\Block{-1}{\cellcolor{orange!10}11 \\ \cellcolor{orange!10}(22\%)} &	\Block{-1}{\cellcolor{blue!5}50 \\ \cellcolor{orange!10}(100\%)} \\

\end{NiceTabular}
\caption{Summary of split grades with respect to PTechs and PTacs, per email type, where 
the abbreviations are {\sf Visual Deception} {\sf (Vis. Dec.)}, {\sf Incentives \& Motivators} {\sf (I\&M)}, {\sf Impersonation} {\sf (Impersona.)}, {\sf Contextualization} {\sf (Context.)}, {\sf Personalization} {\sf (Personal.)}, {\sf Attention Grabbing} {\sf (Att. Grab.)}, {\tt Threat of Loss} {\tt (Loss)}, {\tt Threat to Identity} {\tt (Identity)}, {\tt Claim to legitimate Authority} {\tt (Legit Auth)}, {\tt Fits and Form} {\tt (F\&F)}.
}
\label{tab:Split-scores-table}
\end{table}

In terms of split grades per email type, Table \ref{tab:Split-scores-table} summarizes the number of split grades per PTech and per PTac. We observe that Phishing emails cause more split grades, including
8 (out of the 11) split PTech grades corresponding to 11 emails,
and 23 (out of the 29) split PTac grades with respect to 28 emails.
Since split grades occur only with some PTechs (i.e., 5 out of 8 PTechs) and PTacs (i.e., 4 out of 7 PTacs), the employment of these PTechs and PTacs in certain emails may have caused these emails to be graded high or low by different graders. It should be noted that the calibration process can reduce the subjectivity of the graders, but may not eliminate it. This also explains why {\tt Familiarity} has the highest occurrence among all PTacs and PTechs; the content of an email may be familiar with one grader but not with another. This difference is also reflected by the fact that an email may be graded high by one grader familiar with the email content, and low by another grader not familiar with it.

\begin{insight}
{\it Malicious emails with content familiar to a recipient have a higher sophistication.}

\end{insight}

\subsubsection{RQ 5: Does sophistication evolve with time?}\label{ss:RQ5}

Figure \ref{fig:Sophistication-evolution} plots the yearly mean 
of the PTech grades and of the PTac grades using the aforementioned 124 emails in years 2006, 2011, 2016, and 2021, without eliminating any outlier (if applicable). 
We make two observations. 

\begin{figure}[htbp!]
\centering 
\includegraphics[width=0.45\textwidth]{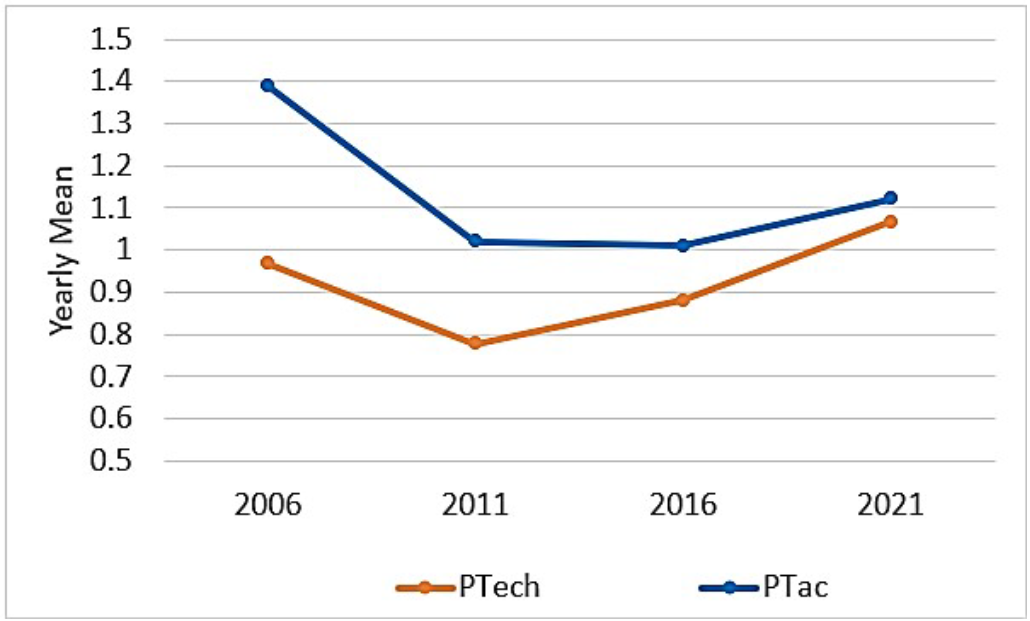}
\caption{\small The average PTech and PTac grades ($y$-axis) in 2006, 2011, 2016, and 2021.
}
\label{fig:Sophistication-evolution} 
\end{figure}

First, email sophistication decreases from 2006 to 2011,
then monotonically increases starting 2011; still, the sophistication in 2006 is higher than that of 2021. 
To see why there is drop in sophistication in 2011, we contrast the 124 emails in 2006 with the 124 emails in 2011 as well as their respective grades. We find that {\it email spoofing} (i.e., a malicious email with a legitimate source address) and {\sf impersonation} (e.g., the attacker impersonates another entity by using its name or logo) were more highly used in 2006 than in 2011. 
Specifically, 70\% of the 124 emails in 2006 alleged to come from banks and financial institutions, while only 21\% of the 124 emails in 2011 alleged to come from banks and financial institutions. 
To see why the sophistication in 2006 is higher than that of 2021, we contrast the 124 emails in 2006 with the 124 emails in 2021 as well as their respective grades. We find that there is an increase in the diversification of impersonating entities. Most impersonating emails in 2006 alleged to come from 3 major institutions (i.e., Banks , PayPal, and Ebay); whereas, 2021 emails have impersonating entities that include banking and financial institutions, as well as delivery companies (e.g., USP, Fedex), Crypto platform (e.g., Coinbase), major stores (e.g., Krogger, Kohl's, Home Depot, Lowe's). The sophistication discrepancy between 2006 and 2021 could also be attributed to that defenses became effective against email spoofing and {\sf Impersonation} around 2011, while these attacks were highly effective prior to 2011. As a result, the attackers turned to new techniques, which may not be as sophisticated as {\sf Impersonation} but are effective, as existing defenses are yet to be effective against them.  

Second, the discrepancy between the PTech grades (i.e., the employment of psychological textual and imagery elements) and the PTac grades (i.e., the attacker's overall deliberate thoughtfulness) 
is significant in 2006 and 2011, but converges over time. This indicates that the attackers have been improving in both aspects over time.

\bigskip
\begin{insight}
{\it Effective defenses against PTechs and PTacs would force attackers to exploit other PTechs and PTacs.
The trend from 2011 to 2021  indicates that attackers always improve their sophistication in both PTechs and PTacs over time.}
\end{insight}

\subsubsection{RQ 6: How do attackers exploit social events?}

We observe that all three email types leverage 
social events.
We observe that Phishing emails leverage both happy events (e.g., Christmas and job offers) and sad events (e.g., flooding and pandemic); Scam emails leverage both happy events (e.g., new vaccine and mortgage refinancing) and sad events (e.g., earthquakes and wars); Spam emails leverage happy events (e.g., buying new homes and holidays sales) and sad events (e.g., health issues such as weight loss and pains). These mean that attackers are \textit{opportunistic attackers}, meaning that they leverage any event that arises to wage attacks regardless of whether it is a tragic event or not.

\begin{figure}[htbp!]
\centering
\includegraphics[width=0.45\textwidth]{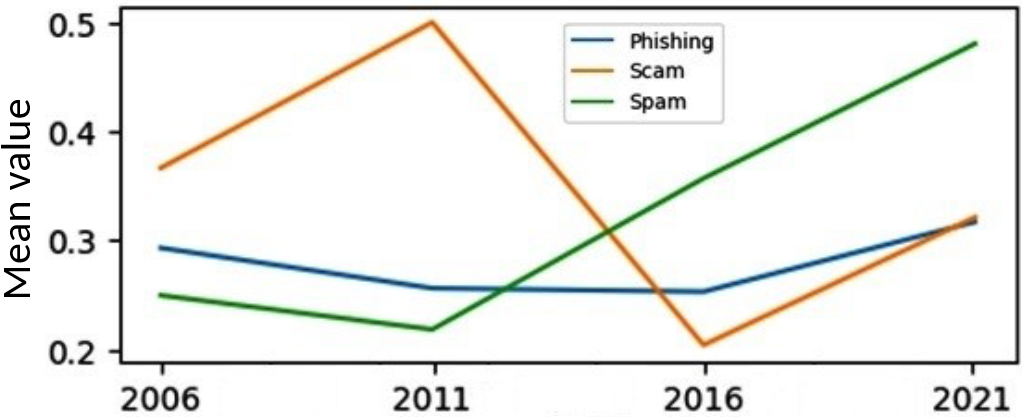}
\caption{{\small The average number of elements (i.e., cues) of the {\sf contextualization}  PTech ($y$-axis) exhibited by the 124 emails in 2006, 2011, 2016, and 2021, which reflects social events. 
}} 
\label{fig:Context-trend} 
\end{figure}

Figure \ref{fig:Context-trend} plots the average {\sf contextualization}, which reflects social events. 
We do not observe any patterns. 
Nevertheless, the significant increase of {\sf contextualization} in Scam emails from 2006 to 2011 can be attributed to how attackers impersonated personalities with exorbitant wealth (e.g., Nigerian Prince, bank manager trying to exfiltrate unclaimed cash). 
Moreover, there seems to be templates for attackers to exploit social events, 
which can be evidenced by the fact that we observe 4 real-world emails that are essentially the same with the only difference that they impersonate 4 different companies (i.e., UPS, T-Mobile, Lowe's, and Home Depot).

\bigskip
\begin{insight}
{\it Social events are widely exploited by attackers in their malicious emails.} 
\label{insight:social-events}
\end{insight}

}

\section{Limitations}
\label{sec:limitations}
The present study has some limitations, which need to be addressed for the future. First, the framework has four limitations: (i) It is based on our understanding of psychological elements (i.e., PTechs and PTacs) that reflect the psychological sophistication of malicious emails. There may be other psychological elements that need to be considered, which can be accommodated by extending our framework. (ii) Its criteria for selecting PTechs and PTacs may not be perfect, meaning that the select PTechs and PTacs may not be complete. Nevertheless, the framework can be trivially extended to accommodate other PTechs and/or PTacs of interest, including those that may be introduced in the future. (iii) The grading rules may need to be refined, to more consistently ensure high levels of consistency in human assessment of email content. (iv) It advocates the use of Kalpha to measure the degree of consistency between graders. However, Kalpha, while widely used, may be less preferred by some researchers. In that case, they can replace the Kalpha with another method for measuring consistency of graders. 

Second, the dataset has four limitations. (i) It may not be representative enough because we only collect and use emails from APWG, though it is arguably the most reputable source in the world. Moreover, even though manually grading 1,036 emails incurs a large amount of work load on each grader, the number of emails is consider small. Future studies need to seek automated grading methods to cope with much larger datasets. 
(ii)  
We admit the potential issue of `informed' graders, as the graders are also the ones that design and revise the grading rules. This may affect the validity of the experimental results to some extent. Nevertheless, the fact that there are still many outliers suggests that the grading process is reliable because there would be no outliers otherwise.
Still, future studies need to separate the group who design grading rules from the group who apply these rules to grade emails.
(iii) While the framework can accommodate any reasonable definitions of PTech and PTac, it would be ideal to assure that the PTechs and PTacs are independent.  For example, the {\sf Urgency} PTech and the {\tt Immediacy} PTac may overlap with each other, and future research needs to revise the PTechs and PTacs to make them independent of each other. (iv) Although the dataset is made up of malicious emails, legitimate emails may also employ PTechs. 
It is interesting to investigate the psychological sophistication of benign emails.


\section{Conclusion}\label{Conclusion}

We have presented a framework for quantifying the psychological sophistication of malicious emails, including Phishing, Spam, and Scam emails. 
The framework is based on two aspects: PTechs and PTacs, both of which are necessary because they respectively reflect the low-level and high-level features of malicious emails. We defined metrics to quantify the sophistication of malicious emails. To measure these metrics, namely to approximate their ground-truth measurements as objectively as possible, we proposed grading rules to guide graders in measuring their sophistication with respect to the PTechs and PTacs. Based on a real-world dataset of 1,036 malicious emails and 4 graders, we draw a number of insights, which deepen our understanding of the sophistication of malicious emails and shed light on how to design effective defenses in the future.  

There are interesting future research directions. (i) It is important to address the limitations of the present study mentioned above. (ii) It is important to extend the framework to cope with other types of cyber social engineering attacks (e.g., messaging-based attacks).
(iii) Having showed that the concept of psychological sophistication is an inherent  feature of malicious emails, it is important to investigate whether defenses, such as malicious email detectors, should be tailored to deal with malicious emails of different degrees of sophistication, or if we should seek  ``one size fits all'' defensive mechanism (i.e., one mechanism that is effective regardless of the degree of sophistication of malicious emails).
(iv) One approach to leveraging the concept of psychological sophistication to guide the design of effective defense is to design a tool that can automatically quantify
the sophistication of an incoming email. This tool would need to be supported by multiple capabilities, such as: (a) automatically recognizing the images contained in emails and then automatically grading their sophistication according to the PTechs; (b) automatically grading the sophistication of incoming emails with respect to PTacs, which would require comprehension of the overall content of an email; and (c) showing that benign emails exhibit different sophistication characteristics (e.g., less sophisticated) than malicious emails, while noting that our framework is equally applicable to quantify the sophistication of benign emails.

\smallskip

\noindent{\bf Acknowledgement}. We thank the APWG for providing most of the emails that are used in the present study. 
Approved for Public Release; Distribution Unlimited. Public Release Case Number 23-3230. The second author is also affiliated  with The MITRE Corporation, which is provided for identification purposes only and is not intended to convey or imply MITRE's concurrence with, or support for, the positions, opinions, or viewpoints expressed by the authors.©2024 The MITRE Corporation. ALL RIGHTS RESERVED.

\bibliographystyle{plain} 
\bibliography{biblio, references, oldlib, rosa_biblio,metrics1_3}
\end{document}